# Survivors in science: A comprehensive analysis of gender disparities

**Marek Kwiek** and **Lukasz Szymula**

UNESCO Chair in Institutional Research and Higher Education Policy, Adam Mickiewicz University, Poznań, Poland

Corresponding Author:
Marek Kwiek
kwiekm@amu.edu.pl

**Keywords:** collaboration gap, gender disparities, global academic career, longitudinal study, publication productivity, women in science

## ABSTRACT

Gender inequality in scientific careers can emerge at two analytically distinct margins: inequality in who remains in science and inequality among those who remain. While the former has been examined in research on attrition in science, much less is known about gender disparities among scientists who remain active for a long period. We address this gap by following “survivors in science”—scientists who started publishing in 2000 and continued publishing until 2020–2023 ($n$ = 41,424). Together, they authored more than two million publications ($n$ = 2,089,097) containing over 70 million cited references ($n$ = 73,118,395) and worked in 38 OECD countries. Using raw Scopus data, we examine gender disparities in publishing intensity, international collaboration, journal prestige, productivity, citation impact, and team size. Our findings reveal that men were on average 23% more productive than women cumulatively in 2000–2023 and 19% more productive in 2019–2023. By contrast, women and men published in similarly prestigious journals, achieved similar field-normalized citation impact, and worked in similarly sized teams. The structure of gender inequality is career-stage dependent: Disparities observed in broader populations of scientists need not persist among those who remain active in science longer, while the gender productivity gap persists.

## 1. INTRODUCTION

Science is a large-scale undertaking, with several million scientists involved in research globally. However, many stop conducting research (in the present conceptualization: stop publishing research results) after 5, 10, or 15 years. Our recent study indicated that about half of all publishing scientists in 38 OECD countries leave science within a decade, and after 19 years only about one third of scientists are still publishing (Kwiek & Szymula, 2025a), with significant gender differences. Attrition in academic science is very high, and it is analyzed in the literature under the “leaving science” theme (Ehrenberg, Kasper, & Rees, 1991; Kaminski & Geisler, 2012; Preston, 2004; Rosser, 2004; Smart, 1990; Spoon, LaBerge et al., 2023; White-Lewis, O’Meara et al., 2023; Xu, 2008; Zhou & Volkwein, 2004). As Ioannidis, Boyack, and Klavans (2014) showed, only a small minority (< 1%) of scientists publish at least one article annually for 16 or more years. This group of about 150,000 scientists contributes significantly to global academic knowledge production with highly cited articles (being responsible for 87.1% of all papers with at least 1,000 citations in the same period).

Gender inequalities in scientific careers can be found in two analytically separate dimensions: unequal probabilities of long-term staying in science (gender inequality in survival) and differences between men and women who stay in science (gender inequality conditional on survival). The first dimension is about selective attrition: Men and women have different probabilities of continuing their scientific careers over time. The second dimension is about whether, after accomplishing specific career milestones and with comparable publishing experience, there still exist gender differences in publishing productivity, international collaboration, journal prestige, citation impact, team formation, and publishing continuity. Conclusions about inequalities in the first dimension (attrition, survival) do not lead to conclusions about inequalities in the second dimension (among survivors): The fact that women leave science earlier and more frequently than men does not determine whether women staying in science for more than two decades still differ from men staying in science for the same period. The present study is focused on the second dimension. Conclusions about gender inequalities among long-term survivors in science are not extrapolated into conclusions about gender inequalities in the whole population of scientists. We ask a longitudinal question: Which gender inequalities remain visible, and which are invisible, among men and women from the same entry cohort who remain active in science (through academic publishing indexed in the Scopus data set) for more than two decades?

This question is complementary to our earlier research questions about attrition and retention in science. We have examined gender differences in leaving science using multiple entry cohorts and Kaplan–Meier survival analysis (Kwiek & Szymula, 2025a). In the present study, survival defines the analytical population. The outcomes we compare are different dimensions of scientific work among those who remained in science. Thus, our previous study examined inequality in survival; the present study examines inequality conditional on survival.

In this study, we follow in a longitudinal fashion consistently productive scientists from the 2000 cohort who were still publishing in 2020–2023 ($n$ = 41,424) and who combined to author over two million articles ($n$ = 2,089,097). We term them “survivors in science”: They are a specific subpopulation of all scientists from OECD countries whose first publication (of any type) in the Scopus data set was dated 2000 and who work in one of 16 disciplines. We examined 13 STEMM and three social science disciplines in which publishing patterns are similar to those in STEMM: BUS (Business, Management, and Accounting), ECON (Economics, Econometrics, and Finance), and PSYCH (Psychology). A relatively small number of long-term survivors is not a sampling deficiency. Long observation windows necessarily produce smaller cohorts because attrition in science accumulates over time. Thus, the size of the survivor population reflects the empirical rarity of sustained long-term scientific careers rather than an attempt to construct a representative sample of all scientists.

We looked at gender differences: We wanted to see to what extent men survivors differ from women survivors in how they published, collaborated, selected academic journals, and were cited over the period. Additionally, we looked at gender differences in the same pool of scientists in 2019–2023 to determine whether gender differences accumulated over a long period differed from recent gender differences. We assumed that there may be differences between cumulative and recent publishing, work, citation, and collaboration patterns.

Our sample consists of successful individuals: They all survived in science for at least 20 years, against the odds, seeing their colleagues leave academic science. They are a minority; the majority of their colleagues left, as we have shown elsewhere (Kwiek & Szymula, 2025a). We studied a specific group of scientists and their publishing, work, and collaboration patterns. The current pool of scientists working in OECD science systems consists of male and female scientists from different cohorts, some working five years and others 15, 25, or more years (in various proportions in different disciplines and countries). This heterogeneous pool of scientists includes individuals from different cohorts. Our study of survivors in science shows gender differences among late-career scientists, not those among young and mid-career scientists. A lack of (or increased) gender differences in this segment of scientists does not reflect a lack of (or increased) gender differences in other populations of interest (i.e., in other academic age groups).

We had a single rationale in this study: Women in science have been considered in much detail across disciplines, countries, and career stages. Gender disparities have been shown in research funding, citations and self-citations, academic hiring and promotions, recognition for teamwork, research collaboration, awards, and much more (Ceci, Kahn, & Williams, 2023; Ductor, Goyal, & Prummer, 2023; King, Bergstrom et al., 2017; Koffi, 2025; Sarsons, 2017; Sarsons, Gërxhani et al., 2020). However, in most cases, the studies were cross-sectional, using bibliometric, survey, interview, case study, national registry, and other data sources. In this research, we follow individuals from the same (academic) age cohort over time, for more than two decades, in a tradition of longitudinal research design (Glenn, 2005; Menard, 2002; Ployhart & Vandenberg, 2010; Singer & Willett, 2003).

Women in science endure gender omission bias and remain underrepresented in math-intensive domains, with implications for review referencing behavior in academic institutions (Koffi, 2025). All-female papers, on average, receive 10% fewer citations than all-male papers—which corresponds well with findings from a gender homophily study of Polish scientists that shows solo-male and all-male papers are published in more prestigious journals than solo-female and all-female papers (Kwiek & Roszka, 2021b). All-male teams omit more papers with female coauthors compared with all-female teams, and vice versa (Koffi, 2025).

Examining men and women in science in general differs fundamentally from examining men and women with exactly the same long-term publishing experience. The subject of survivors has been largely overlooked, as both longitudinal bibliometric data and longitudinal methods (e.g., Kaplan–Meier survival analysis used in exit studies) have been underused. The research design employed for this article relies on an underlying assumption that despite their limitations, large, curated bibliometric data sets can extend our knowledge of individuals and groups in science from a temporal perspective (Kwiek & Szymula, 2025a, b). The "survivors in science" are scientists fortunate enough to have avoided exit from science and able to show their research abilities in the same, long period. What we learn about them in this research is what we learn about late-career scientists; while constituting a minority, they extend our insights about the science profession. In this study, we follow a single cohort (2000), but we could have also followed other cohorts (e.g., 2005 and 2010) until 2023. What can be learned from longitudinal studies of cohort(s) cannot be learned from cross-sectional studies in which scientists from various cohorts are mixed with one another.

Our unique data set allowed us to closely analyze a clearly defined sample. We were able to compare men and women in the same disciplines with the same publishing experience, active in academic science for up to 24 years (i.e., still publishing in academic journals). The scale of this research allowed us to compare men and women in a number of fundamental dimensions of academic work: lifetime scholarly output, annual publication productivity, intensity of international research collaboration, journal selection patterns, being cited, team formation, and publishing frequency. However, we were not able to examine the "work climates" characterizing the basic units in which scientists work, which are reported to be especially important in STEM disciplines for productivity (Fox & Mohapatra, 2007).

We were interested in the following set of dimensions: gender differences in publishing intensity (lifetime scholarly output); gender differences in international research collaboration intensity; gender differences in journal publishing patterns (Scopus journal percentile ranks); gender differences in publishing productivity patterns; gender differences in being cited (four-year field-weighted citation impact—FWCI 4y); gender differences in team formation; and gender differences in yearly publishing patterns/publishing breaks.

Long-term survivors in science are an especially useful population for such comparisons. Their defining feature is persistent presence in science (rather than their membership in research performance elite). A long-term presence in science should not be identified with membership in the upper echelons of productivity or citation distribution. Survivors include both low and high research performers. The selective nature of this population derives from publishing for more than two decades, with many colleagues from their original cohort leaving science in the meantime. After two decades of attrition, we observe only those who remain in science (who keep publishing) and ask which of the well-documented gender inequalities are still visible in this population.

Long-term survivors have not been discussed in studies of gender inequalities in science. Reasons for this gap include data and methodology: Large, longitudinal bibliometric data sets and longitudinal methods have been used in examining academic careers much less often than cross-sectional data sets and approaches. Currently, curated bibliometric data and the availability of computer power allow researchers to follow scientists from their first indexed publications for long periods. Academic careers can be studied from a temporal perspective (Kwiek & Szymula, 2025a, b). What can be learned from a cohort followed for more than two decades cannot be learned from cross-sectional studies in which scientists from different entry cohorts and with different lengths of academic experience are mixed.

We analyze gender differences in seven dimensions: lifetime scholarly output; international research collaboration; journal publishing patterns measured through Scopus journal percentile ranks; publication productivity; citation impact measured through FWCI 4y; team formation; and publishing continuity and publishing breaks. These dimensions allow us to examine whether gender disparities documented for broader populations of scientists are also present among women and men who have remained active in science for more than two decades.

## 2. CONCEPTUAL FRAMEWORK

This research contributes to the literature on the academic careers of and differences between men and women in science, specifically late-career scientists. It belongs to a pool of large-scale bibliometric studies of a longitudinal (Menard, 2002; Ployhart & Vandenberg, 2010; Singer & Willett, 2003) and cohort (Glenn, 2005) nature, i.e., to studies where individual scientists belong to the same academic age cohort and are followed over time, from their first publication onward. Most research on women in science is neither cohort-based nor longitudinal, with samples mixing scientists of different biological ages (or academic ages) and disregarding their publication progression from the first publication onward (Huang, Gates et al., 2020; King et al., 2017; Larivière, Ni et al., 2013; Morgan, Way et al., 2021). We were interested in the cumulative effects of various working and collaboration habits over the last two decades.

We could have selected cumulative effects after any number of years, but cohort 2000 provided us enough individuals (and their micro-data), and the last two decades of accumulation proved long enough. The longer the period studied, the smaller the number of scientists due to high attrition, especially of women. Explanations for "leaving science" or "quitting science" include problems with work–life balance (Rosser, 2004; Smart, 1990), low salaries compared with other professionals and low job security (Zhou & Volkwein, 2004), colleague concerns and workload concerns (Wohrer, 2014), various types of discrimination in the workplace (Preston, 2004; Smart, 1990), and chilly work climates for women (Cornelius, Constantinople, & Gray, 1988; Spoon et al., 2023).

Various gender gaps are discussed in the literature on women in science (Kwiek & Roszka, 2022b). The two most often examined are the gender productivity gap and the gender collaboration gap. There is a long tradition, originating in the 1970s (Cole, 1979), of literature on women and publishing productivity which tends to show that women are less productive than men. One strand also indicates that women are less productive early in their careers when they have more family obligations, but later in their careers, they are able to catch up with men, with the gender productivity gap disappearing. The literature shows gender productivity gaps in different countries, periods, and STEMM disciplines (Abramo, D'Angelo, & Di Costa, 2019; Fox & Mohapatra, 2007; Fox & Nikivincze, 2021; Larivière, Vignola-Gagné et al., 2011; Nielsen, 2016; van den Besselaar & Sandstrom, 2016). Ceci et al. (2023) show in their overview of gender biases in science that both time and productivity distribution matter: Gender differences in productivity vary between graduate studies, tenure-track hiring, and tenure; and they also vary between the upper tail of the distribution and the remaining levels of productivity (with the largest differences being observed for the top 10% of the productivity distribution). Ceci et al. (2023) also rightly note that findings on productivity rely heavily on definitions of their populations.

There is also a long tradition of literature on women and research collaboration, and especially international research collaboration, which generally suggests that while women are generally more collaborative (show a higher inclination to be involved in research collaboration in general), they are less inclined to participate in international research collaboration. This literature indicates the role of this gender collaboration gap in academic progression and promotion and in research productivity (Abramo, D'Angelo, & Murgia, 2013; Aksnes, Piro, & Rørstad, 2019; Bozeman, Fay, & Slade, 2012; Fox, Realff et al., 2017; Larivière, Larivière, & Gingras, 2013; Maddi et al., 2019; Ductor et al., 2023). As Ductor et al. (2023) show, women tend to work more with the same coauthors; women coauthor more with more experienced authors at each stage of their career; and women choose collaboration patterns that are correlated with lower output.

The literature also shows a gender mobility gap (Ackers, 2008; Frehill & Zippel, 2010; Uhly, Visser, & Zippel, 2017) which indicates that women in science are less involved in (especially international) physical mobility than men due to family and caregiving obligations. Female scientists are also reported to cite themselves less often than male scientists, which is termed the gender self-citation gap (Hutson, 2006; King et al., 2017; Maliniak, Powers, & Walter, 2013; Mishra, Fegley et al., 2018). Fewer self-citations lead to fewer external citations (Fowler & Aksnes, 2007). And female scientists are less often awarded research grants, and their grants are smaller, which is termed the gender research funding gap (Cruz-Castro & Sanz-Menendez, 2019; van den Besselaar & Sandstrom, 2015).

There is also literature on women and journal selection which tends to show that women are less inclined to submit their manuscripts to top journals compared to men and that women generally submit their manuscripts to less prestigious journals (Mihaljević-Brandt, Santamaria, & Tullney, 2016; Sugimoto & Larivière, 2023). Women are also less cited than men, with implications for academic careers and especially for full professorships—which is termed the gender citation gap (Abramo, D'Angelo, & Rosati, 2015; Aksnes, Rørstad et al., 2011; Ghiasi, Mongeon et al., 2018; Huang et al., 2020; Lerchemueller, Hoisl, & Schmallenbach, 2019; Maddi et al., 2019; Madison & Fahlman, 2020; Maliniak et al., 2013; Potthof & Zimmermann, 2017; Thelwall, 2020; van den Besselaar & Sandstrom, 2017).

There is also a gender professional network gap: Women have narrower and less international formal (and informal) collaboration networks than those of men (Clauset, Arbesman, & Larremore, 2015; Feeney & Bernal, 2010; Halevi, 2019; Kegen, 2013; van den Brink & Benschop, 2013). The literature also reports a gender academic time distribution gap (Cummings & Finkelstein, 2012; Goastellec & Vaira, 2017; Toutkoushian & Bellas, 1999) and a gender academic role orientation gap (Leisyte & Hosch-Dayican, 2017; Miller & Chamberlin, 2000): Women spend less time on research and more on teaching compared with men, and women are less research-oriented and more teaching-oriented. These two gaps are studied in international comparative academic profession surveys that allow comparisons of weekly working time distribution and academic role orientation.

There are four other gender gaps discussed in the literature worth noting. There is a gender methods gap, in which women use quantitative methods less often and qualitative methods more often, with implications for the academic journal selection. More quantitatively focused journals tend to be more prestigious (Key & Summer, 2019; Thelwall, Bailey et al., 2019). There is a gender group work recognition gap, which means that women receive less deserved recognition or less deserved credit for their collaborative publications than men do (Heffner, 1979; Sarsons, 2017; Sarsons et al., 2020). In other words, collaborative papers tend to support the academic careers of men more strongly than the academic careers of women. Finally, the gender tenure gap means that women are less often promoted to tenure despite equal achievements (Diezmann & Grieshaber, 2019; Fell & Konig, 2016; Rivera, 2017; Weisshaar, 2017), and the gender salary gap means that women have lower salaries in the same positions as men (Barbezat & Hughes, 2005; Ceci, Ginther et al., 2014; Ward & Sloane, 2000).

All these disparities, conceptualized as gender gaps, are valid for women scientists in general, in various countries and disciplines, mostly regardless of their biological or academic age (academic experience). Most studies on women in science are case studies: Their findings generally come from small-scale interview and survey research conducted in Anglophone countries in the past two decades. None are cohort studies, and very few are longitudinal in the strict sense of the term (Menard, 2002, pp. 2–3). The conclusions from the numerous small-scale studies paint the general picture summarized above.

This study, in contrast, is large in scale and longitudinal, bibliometric, and quantitative in nature. We were able to compare the working and collaboration habits of men and women in 16 disciplines in the 38 OECD countries. We compared men and women in the same period within the same disciplines. Our study differs from that of Ioannidis, Boyack, & Klavans (2014) in being cohort-based and longitudinal; their "continuously publishing core" of 150,000 scientists were members of numerous cohorts, as opposed to our survivors in science who originated from a single cohort. Ioannidis et al. examined stability in science, with gender ignored; our focus was on gender disparities in how late-career scientists publish, collaborate, and are cited. Additionally, Ioannidis et al. were interested in comparing the core with the noncore, or scientists who were not continuously publishing; in contrast, our focus was on comparing men and women among survivors in science. A comparison with nonsurvivors would refer to the leavers from science, which would involve attrition analysis (our topic elsewhere: Kwiek & Szymula, 2025a).

Here, we were interested in whether selected gender gaps held for survivors in science. We focused on the following: the productivity gap; the collaboration gap; the publishing pattern gap; the citation gap; and two rarely studied gaps: the team formation gap and the publishing frequency gap.

## 3. DATA AND METHODS

Our research design omits the phenomenon of exit from science, extremely important in understanding gender inequality in science, as we have discussed this research theme in much detail in an earlier study in which leaving science in 38 OECD countries in 16 STEM disciplines was empirically based on two major cohorts: the cohort 2000 ($n$ = 142,776) and the cohort 2010 ($n$ = 232,843), and nine cohorts between 2000 and 2010 (Kwiek & Szymula, 2025a). In that study design, scientists were followed from the year 2000 (and 2010) until 2022 or until they left. Leaving science was conceptualized as quitting publishing in scientific journals. We used survival analysis to compare attrition of men and women over time.

In the current research we start where the previous research was stopped: On average across 16 STEM disciplines, the Kaplan–Meier probability of staying in science is 29.4% (95% confidence interval: 29.0–29.8%) for women and 33.6% (95% confidence interval: 33.3–33.9%) for men after 19 years of publishing experience. Consequently, our current focus is on those who started publishing in 2000 and continued publishing until 2020–2023 (here our observation window is longer). Our data set is particularly well suited for longitudinal studies and the novelty of this research is to examine only those successful few "survivors in science," without claiming relevance to refer to other segments of science. We have deliberately excluded those who left science as they have been carefully examined previously in our exit research (Kwiek & Szymula, 2025a).

To enter the pool of our survivors in science, the scientists needed to have at least 10 publications (journal articles or conference proceedings in journals or books) in their individual publishing portfolio and at least one publication in 2020–2023 (in a four-year window). We observed a homogeneous pool of scientists, men and women, who started publishing in the same year, under similar structural conditions, with similar publishing and collaboration pressures and comparable research funding opportunities.

### 3.1. How Was the Sample Obtained?

The study utilized a Scopus database snapshot from March 29, 2024. Based on author identifiers (initially $n$ = 49,554,504 authors in the Scopus data set), the first and last publication year was determined for each author. Author-level metrics were then calculated based on their individual publication portfolios (articles in journals or conference proceedings published in journals or books only; restricted to no later than 2023). For each author, the following metrics were determined: scholarly output, international collaboration rate, average journal percentile rank, productivity normalized by journal prestige, average FWCI 4y, median team size, and median publishing break. Authors with no scholarly output for the selected portfolio were removed from the sample ($n$ = 41,210,209; $n_{removed}$ = 8,344,295). Then each author was subsequently assigned to a country, gender, and discipline. The author set was restricted to scientists with affiliation from one of the OECD countries ($n$ = 20,964,249; $n_{removed}$ = 20,245,960), a defined gender ($n$ = 14,790,196; $n_{removed}$ = 6,174,053) and a selected STEMM or three SOC disciplines ($n$ = 12,341,405; $n_{removed}$ = 2,448,791). The author set was then limited to those with at least 10 publications in their portfolios, whose first publication year was 2000 and whose last publication year fell between 2020 and 2023 inclusive (final sample $n$ = 41,424).

To create the sample, the bibliometric database Scopus, made available by ICSR Lab, was used. Access to the database and the ability to perform cloud-based computations were provided through the Databricks platform. PySpark notebooks were developed and executed on the full Scopus database using a cluster in standard mode with Databricks Runtime version 11.2 ML, Apache Spark technology version 3.3.0, Scala 2.12, and an i3.2xlarge instance with 61 Gbyte of memory, eight cores, and one to six workers for the worker type, as well as a c4.2xlarge instance with 15 Gbyte of memory and four cores for the driver type.

The definition of "survival" needs a detailed explanation. The criterion used (publishing in 2000, having at least 10 publications in the publication portfolio, publishing in 2020–2023) is based on our examination of median publishing breaks for the 2000 cohort of scientists by discipline and gender, used for the first time in our study on quantifying attrition in science (Kwiek & Szymula 2025a, Supplementary material) and repeated in this research for a different time span. In the exit study, the percentage of scientists from the 2000 cohort who published a paper every year (publishing break: zero years) was 86.70% for men and 83.66% for women (all disciplines combined), and it was about 2% (1.88% and 2.26%, respectively) for those having a publishing break of four years. In this research, as Table 11 in Section 4.7 shows, the publishing break of four years can be attributed to 0.00% and 0.01% of scientists in social sciences (women vs. men) and to 0.10% and 0.07% of scientists in STEM disciplines (women vs. men). For instance, in seven disciplines for women and in eight disciplines for men (out of 16), the median publishing break of four years or more was attributed to 0.00% of scientists in the sample. Consequently, our selection of the sample, using a four-year window (2020–2023) at the end of the publishing bracket, is well designed to capture all or almost all (depending on the discipline) publishing scientists from the cohort 2000.

There are widely discussed differences in publication practices between the disciplines examined in this research (e.g., business and economics as comparatively low producers and physics as a large producer); however, our publishing threshold of four years (2020–2023) allows us to consider scientists with both a slower and a faster publishing pace.

It needs to be added that all gender differences are shown within disciplines (women in economics vs. men in economics etc.), with no cross-disciplinary comparisons being made. It is especially important for us not to draw cross-disciplinary conclusions as different disciplines have different publication paces and different production cycles. The sample does not disproportionately capture certain types of researchers or subfields and does not exclude researchers publishing at a slower frequency (as we have shown for the four-year publication window of 2020–2023). The sample includes all or almost all scientists from the cohort 2000 publishing for more than two decades, whose gender, discipline, and country of affiliation (OECD) could be determined.

### 3.2. How Was Gender Defined?

Determining gender in the sample relied on a gender data set provided by the ICSR Lab platform. The data set was filtered to scientists with a defined gender as man or woman only with a probability score greater than or equal to 0.85. Gender assignment was based on Elsevier's solution, which embedded the Namsor gender determination tool. Gender classification depended on three input features: the author's first name, last name, and country of origin. Country was defined as the author's dominant country from the year of their first publication based on Scopus data. If the author had more than one dominant country, then no value was assigned. The Namsor tool returned both predicted gender and probability score (Elsevier, 2020, pp. 122–123). We decided to use Scopus rather than Web of Science data because our unit of analysis was the individual scientist (rather than the individual publication); and "in terms of author disambiguation, Scopus is more accurate" (Sugimoto & Larivière, 2018, p. 36).

### 3.3. How Were the Disciplines Defined?

To identify the dominant discipline of the scientists, a Scopus data set, including publications published up to and including 2023, was analyzed. From this data set, specific columns were selected: publication identifiers, author identifiers, and cited references. Each cited reference was associated with at least one discipline, based on the classification of the journal in which the cited reference appeared. Disciplines were initially defined using the four-digit All Science Journal Classification (ASJC) codes provided by Scopus. Then, only the first two digits of the ASJC codes were used, resulting in broader, two-digit ASJC categories. For each author, the number of cited references across different disciplines was counted, with the "multidisciplinary" category excluded from this process. The dominant discipline for a given author was defined as the discipline associated with the highest number of cited references (i.e., the modal discipline). This resulted in a table mapping each author's identifier to their dominant discipline. In cases where an author did not have a clearly dominant discipline or had multiple disciplines tied for the highest count, the author was excluded from the data set. Finally, the data set was further filtered to retain only those authors whose dominant discipline belonged to either the STEMM or SOC group. To determine the dominant discipline of our sample, we used more than 70 million cited references from their publications ($n$ = 73,118,395), on average 35 cited references per publication. Thus, the dominant discipline was not based on journals in which a scientist published most frequently (and their disciplinary locations), but on the modal discipline of the literature cited across the scientist's entire publication portfolio (in all their indexed publications).

### 3.4. How Were the Countries Defined?

To determine the dominant country of scientists, a Scopus data set of publications from 2023 and earlier was used. From this data set, columns containing publication identifiers, author identifiers, and the countries associated with each author were selected. For each author, the number of times each country was listed in all of their publications was counted. The country with the highest frequency (i.e., the modal value) was identified as the author's dominant country. Authors who had multiple countries tied for the highest count or no assigned country were excluded from the data set. Finally, the table was filtered to include only scientists affiliated with an OECD country (Kwiek & Szymula, 2023). We could have used a more fine-grained approach to country assignment; for instance, we could have applied dominant countries to individuals in specific periods of time since 2000 (2000–2004, 2005–2009, 2010–2014 etc.) to make sure that their international migration patterns become visible. However, the focus of this study is on individuals with their gender defined and who were embedded in disciplines. We do not study cross-national differences; country-related differences could be between OECD countries (our pool of scientists) and non-OECD countries, a comparison that we do not do.

### 3.5. How Were Author-Level Metrics Calculated?

The assignment of author-level metrics was based on indicators derived from each author's publication portfolio (articles in journals and conference proceedings published in journals or books only). First, for every publication in the portfolio, the team size was determined as the total number of authors of the publication, and the number of unique countries was calculated based on all author affiliations listed in the publication. These two values enabled a binary classification of whether a publication was collaborative ( = 1 if the publication included more than two authors) and whether it was international ( = 1 if at least two unique countries were represented among the affiliations and the publication was collaborative). Each publication record was also supplemented with the percentile rank of the journal for 2023 and FWCI 4y value. Both indicators were available on the ICSR Lab platform. In cases where the journal percentile rank was missing, a value of 1 was assigned; in cases where the FWCI value was missing, a value of 0 was used. Subsequently, the data were aggregated to compute the following author-level metrics: scholarly output; international collaboration rate (calculated as the ratio of all international publications to all collaborative publications); average journal percentile rank; productivity normalized by journal prestige (sum of journal percentile ranks for publications divided by 24 years); average FWCI 4y; median team size; and median publishing break.

### 3.6. Sample Description

The characteristics of the sample are shown in Table 1. We studied $n$ = 41,424 scientists (of which 33.86% were women), authors of more than two million articles ($n$ = 2,089,097). The disciplines with the most survivors were MED and BIO (40.94% and 14.79% of our sample, respectively) and the countries most represented were the United States and Japan (27.22% and 9.14%, respectively). The three disciplines with the highest percentage of women survivors in science were IMMU, MED, and BIO (46.79%, 41.13%, and 38.60%, respectively) and the three countries with the highest percentage of women survivors in science were Portugal, Italy, and Poland (51.95%, 45.97%, and 45.68%, respectively). The three social science disciplines represented in the sample varied in their gender composition, with ECON having 20.03% women, BUS 33.33% and PSYCH 47.93%. We selected these three social science disciplines because scientists show publishing habits similar to those of STEMM disciplines, especially in publishing in journals and attaching significance to journal type (e.g., the role of publishing in English and the role of top journals in promotions and academic careers). The lowest shares of women were seen in ENG (11.08%) and PHYS (16.28%).

The ASJC discipline codes used in this research were as follows: AGRI: Agricultural and Biological Sciences; BIO: Biochemistry, Genetics, and Molecular Biology; BUS: Business, Management, and Accounting; CHEM: Chemistry; COMP: Computer Science; EARTH: Earth and Planetary Sciences; ECON: Economics, Econometrics, and Finance; ENG: Engineering; ENVIR: Environmental Science; IMMU: Immunology and Microbiology; MATER: Materials Science; MATH: Mathematics; MED: Medicine; NEURO: Neuroscience; PHYS: Physics and Astronomy; PSYCH: Psychology. The three non-STEMM disciplines were BUS, ECON, and PSYCH.

## 4. RESULTS

### 4.1. Publishing Patterns—Lifetime Scholarly Output

The composition of our pool of survivor scientists gave us an opportunity to compare the lifetime scholarly output (cumulative) of men and women over 20–24 years in 16 disciplines. In 12 disciplines, the cumulative output of men in 2023 was higher than the cumulative output of women, and the difference is statistically significant. Graphically, the difference can be seen in Figure 1, which presents kernel density plots for each discipline. Women in the plots are often located more in the lower parts of the output distribution, and men are often located in the upper parts. This is especially visible in BIO, IMMU, and MATH.

**Table 1.** Sample: all scientists from the 2000 cohort who survived until 2020–2023 (who were still publishing in 2020–2023). "Other" represents 18 smaller OECD science systems

| | | ***n* women** | **% col** | **% row** | ***n* men** | **% col** | **% row** | ***n* total** | **% col** |
|---|---|---|---|---|---|---|---|---|---|
| | TOTAL | 14,027 | 100 | 33.86 | 27,397 | 100 | 66.14 | 41,424 | 100 |
| | TOTAL SOCIAL | 658 | 4.69 | 34.56 | 1,246 | 4.55 | 65.44 | 1,904 | 4.60 |
| | TOTAL STEMM | 13,369 | 95.31 | 33.83 | 26,151 | 95.45 | 66.17 | 39,520 | 95.40 |
| Discipline | AGRI | 1,038 | 7.40 | 34.06 | 2,010 | 7.34 | 65.94 | 3,048 | 7.36 |
| | BIO | 2,365 | 16.86 | 38.60 | 3,762 | 13.73 | 61.40 | 6,127 | 14.79 |
| | BUS | 187 | 1.33 | 33.33 | 374 | 1.37 | 66.67 | 561 | 1.35 |
| | CHEM | 535 | 3.81 | 30.11 | 1,242 | 4.53 | 69.89 | 1,777 | 4.29 |
| | COMP | 247 | 1.76 | 18.47 | 1,090 | 3.98 | 81.53 | 1,337 | 3.23 |
| | EARTH | 384 | 2.74 | 25.00 | 1,152 | 4.20 | 75.00 | 1,536 | 3.71 |
| | ECON | 124 | 0.88 | 20.03 | 495 | 1.81 | 79.97 | 619 | 1.49 |
| | ENG | 222 | 1.58 | 11.08 | 1,782 | 6.50 | 88.92 | 2,004 | 4.84 |
| | ENVIR | 296 | 2.11 | 30.27 | 682 | 2.49 | 69.73 | 978 | 2.36 |
| | IMMU | 175 | 1.25 | 46.79 | 199 | 0.73 | 53.21 | 374 | 0.90 |
| | MATER | 182 | 1.30 | 22.75 | 618 | 2.26 | 77.25 | 800 | 1.93 |
| | MATH | 158 | 1.13 | 21.29 | 584 | 2.13 | 78.71 | 742 | 1.79 |
| | MED | 6,975 | 49.73 | 41.13 | 9,982 | 36.43 | 58.87 | 16,957 | 40.94 |
| | NEURO | 299 | 2.13 | 36.87 | 512 | 1.87 | 63.13 | 811 | 1.96 |
| | PHYS | 493 | 3.51 | 16.28 | 2,536 | 9.26 | 83.72 | 3,029 | 7.31 |
| | PSYCH | 347 | 2.47 | 47.93 | 377 | 1.38 | 52.07 | 724 | 1.75 |
| Country | United States | 2,062 | 14.70 | 18.28 | 7,154 | 26.11 | 81.72 | 11,277 | 27.22 |
| | Japan | 590 | 4.21 | 15.58 | 3,197 | 11.67 | 84.42 | 3,787 | 9.14 |
| | United Kingdom | 1,066 | 7.60 | 35.75 | 1,916 | 6.99 | 64.25 | 2,982 | 7.20 |
| | Italy | 1,311 | 9.35 | 45.97 | 1,541 | 5.62 | 54.03 | 2,852 | 6.88 |
| | Germany | 644 | 4.59 | 22.84 | 2,176 | 7.94 | 77.16 | 2,820 | 6.81 |
| | France | 979 | 6.98 | 35.18 | 1,804 | 6.58 | 64.82 | 2,783 | 6.72 |
| | Spain | 661 | 4.71 | 40.33 | 978 | 3.57 | 59.67 | 1,639 | 3.96 |
| | Canada | 591 | 4.21 | 38.30 | 952 | 3.47 | 61.70 | 1,543 | 3.72 |
| | Australia | 545 | 3.89 | 40.28 | 808 | 2.95 | 59.72 | 1,353 | 3.27 |
| | South Korea | 188 | 1.34 | 14.52 | 1,107 | 4.04 | 85.48 | 1,295 | 3.13 |
| | Netherlands | 365 | 2.60 | 38.38 | 586 | 2.14 | 61.62 | 951 | 2.30 |
| | Poland | 407 | 2.90 | 45.68 | 484 | 1.77 | 54.32 | 891 | 2.15 |

**Table 1.** (*continued*)

| | *n* women | % col | % row | *n* men | % col | % row | *n* total | % col |
|---|---|---|---|---|---|---|---|---|
| Turkey | 232 | 1.65 | 30.77 | 522 | 1.91 | 69.23 | 754 | 1.82 |
| Switzerland | 173 | 1.23 | 28.04 | 444 | 1.62 | 71.96 | 617 | 1.49 |
| Sweden | 241 | 1.72 | 39.31 | 372 | 1.36 | 60.69 | 613 | 1.48 |
| Greece | 160 | 1.14 | 32.92 | 326 | 1.19 | 67.08 | 486 | 1.17 |
| Belgium | 160 | 1.14 | 35.56 | 290 | 1.06 | 64.44 | 450 | 1.09 |
| Mexico | 149 | 1.06 | 33.48 | 296 | 1.08 | 66.52 | 445 | 1.07 |
| Israel | 163 | 1.16 | 37.13 | 276 | 1.01 | 62.87 | 439 | 1.06 |
| Portugal | 226 | 1.61 | 51.95 | 209 | 0.76 | 48.05 | 435 | 1.05 |
| Other | 1,053 | 7.51 | 34.96 | 1,959 | 7.15 | 65.04 | 3,012 | 7.27 |

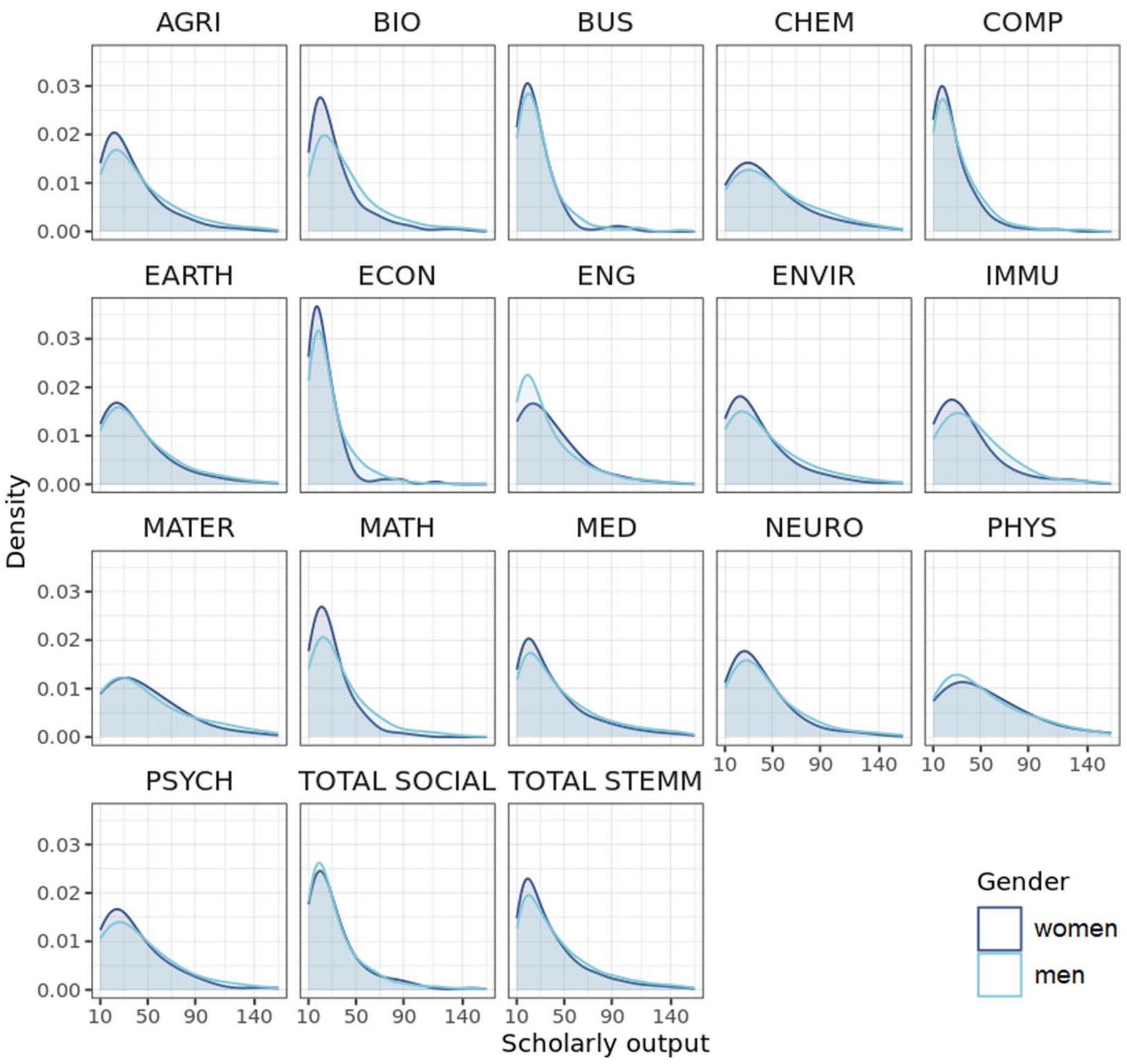

**Figure 1.** Distribution of lifetime scholarly output, all publications for 2000–2023; kernel density plot by discipline and gender.

We will discuss primarily the tables; however, we will also refer to kernel density plots. We use them to visualize the underlying distribution of the data (as in Figure 1). The advantage of density plots over histograms is that they are better interpretable for 18 distributions of our disciplinary subsamples from a gender perspective (all disciplines, all social science disciplines combined, all STEMM disciplines combined). The continuous curve is estimated from the data using kernel density estimation and a Gaussian kernel. Density plots may not work to visualize small data sets, but they tend to be reliable for large data sets like ours. Kernel density plots are used to identify the shape of the distribution (e.g., whether the distribution is symmetric or skewed, with one or more peaks). Density plots offer a continuous representation of the data distribution, with no jagged appearance characteristic of histograms. The area under the curve always adds up to 100%. The smoothness enhances the visual interpretation of the underlying patterns in the data. Density curves allow a quick understanding of the distribution of values in our (disciplinary) data sets.

The statistical details of the average lifetime scholarly output (2000–2023) by discipline and gender are provided in Table 2. The table shows statistics and MFR rate: average male output divided by average female output. An MFR rate higher than 1 means that average output for men is higher than the average output for women. The difference between men and women in lifetime output for all STEMM disciplines was 24% (women: 43.60 publications, men 53.93 publications, MFR = 1.24), and it is statistically significant ($p < 0.001$); for all social science disciplines, in contrast, the difference is not statistically significant. The largest differences were observed for IMMU, MATH, and BIO and were about 30% in favor of men (MFR = 1.35, $p < 0.01$; MFR = 1.33, $p < 0.001$ and MFR = 1.30, $p < 0.001$, respectively). A difference of about 20% in favor of men was observed for AGRI (MFR = 1.24, $p < 0.001$), ENVIR (MFR = 1.25, $p < 0.001$), MATER (MFR = 1.22, $p < 0.01$), and MED (MFR = 1.24, $p < 0.001$). In the social science disciplines, the difference was statistically significant in two out of three (ECON MFR = 1.18, $p < 0.05$ and PSYCH MFR = 1.21, $p < 0.01$).

In almost all disciplines, the gender gap in lifetime scholarly output was substantial; in three, the gap reached about one third and in eight reached about one fifth in favor of men. Percentages translate into publication numbers; for instance, in BIO, men's lifetime scholarly output was on average 44, and women's was on average about 34; and in MATH, it was on average about 38 for men as opposed to on average about 28 for women. Differences in publication numbers can translate into faster promotion to higher academic ranks; however, we were not able to see how publication quantity (and quality) translated into academic promotions using our data set.

We were interested whether the cumulative gender gap in lifetime scholarly output could be caused by men publishing more than women in early and mid-career periods when women potentially have more children and family obligations compared to men. We studied the last five years in the data set to see whether publication numbers for men and women were on average more equal, following early, more substantial differences in scholarly output (Table 3). However, in 2019–2023, the gender patterns of average scholarly output were almost exactly the same as the cumulative patterns for 24 years. In all STEMM disciplines combined, men published on average 24% more articles than women (MFR = 1.24, $p < 0.001$), and for the three disciplines with the highest gender difference, the gap was about one third in favor of men (BIO MFR = 1.30, $p < 0.001$, MATH MFR = 1.33, $p < 0.001$, and IMMU MFR = 1.35, $p < 0.01$). The hypothesis that the gender productivity gap is rooted in early career periods when women have more time-consuming children and family responsibilities and disappears later had to be rejected. The kernel density plots in Figure 1 also clearly show that in all disciplines, the majority of observations are in the 10–50 range of publications (10 publications being the threshold to enter the subpopulation studied) and the number of top performers, with 100 or more papers, is very small. MATH and COMP had the largest share of scientists with lower numbers of publications.

**Table 2.** Average scholarly output, all publications for 2000–2023, by discipline and gender

| | Average | | Std | | | | |
|---|---|---|---|---|---|---|---|
| **Discipline** | **Women** | **Men** | **Women** | **Men** | **MFR** | ***Z*-statistic** | ***p*-value** |
| AGRI | 37.35 | 46.21 | 27.26 | 38.86 | 1.24 | −7.31 | <0.001 |
| BIO | 33.84 | 44.01 | 25.90 | 36.82 | 1.30 | −12.67 | <0.001 |
| BUS | 25.91 | 29.03 | 16.55 | 20.50 | 1.12 | −1.94 | 0.052 |
| CHEM | 51.57 | 60.37 | 50.38 | 54.42 | 1.17 | −3.30 | <0.001 |
| COMP | 27.47 | 32.17 | 29.05 | 33.29 | 1.17 | −1.28 | 0.201 |
| EARTH | 39.37 | 46.60 | 29.76 | 42.07 | 1.18 | −3.69 | <0.001 |
| ECON | 23.82 | 28.07 | 16.18 | 20.97 | 1.18 | −2.45 | <0.05 |
| ENG | 36.69 | 37.74 | 29.42 | 36.77 | 1.03 | −0.49 | 0.624 |
| ENVIR | 37.28 | 46.70 | 28.19 | 40.47 | 1.25 | −4.18 | <0.001 |
| IMMU | 37.69 | 50.73 | 27.96 | 54.24 | 1.35 | −2.97 | <0.01 |
| MATER | 49.46 | 60.22 | 42.18 | 57.81 | 1.22 | −2.76 | <0.01 |
| MATH | 28.20 | 37.53 | 16.43 | 30.99 | 1.33 | −5.10 | <0.001 |
| MED | 45.38 | 56.15 | 42.89 | 57.25 | 1.24 | −14.00 | <0.001 |
| NEURO | 40.85 | 46.85 | 31.05 | 38.31 | 1.15 | −2.43 | <0.05 |
| PHYS | 94.72 | 92.80 | 168.82 | 177.68 | 0.98 | 0.23 | 0.818 |
| PSYCH | 40.07 | 48.67 | 33.09 | 43.93 | 1.21 | −2.99 | <0.01 |
| TOTAL SOCIAL | 32.98 | 34.59 | 27.56 | 31.14 | 1.05 | −1.16 | 0.246 |
| TOTAL STEMM | 43.60 | 53.93 | 50.49 | 73.43 | 1.24 | −16.40 | <0.001 |

Lifetime scholarly output is a very crude measure of academic work: No distinction is made regarding journal type (i.e., all journals are counted equally). We examine more fine-grained productivity below in Sections 4.3 and 4.4.

### 4.2. International Research Collaboration

As high research productivity is generally linked with international collaboration, we were interested in gender differences in the international collaboration rate. The international collaboration rate for 2000–2023 was calculated as all collaborative journal articles or conference proceedings in journals or books with international coauthors (an international affiliation in the publication's byline) divided by all collaborative journal articles or conference proceedings in journals or books (of any type: international, national, institutional).

**Table 3.** Average scholarly output, all publications for 2019–2023, by discipline and gender

| | Average | | Std | | | | |
|---|---|---|---|---|---|---|---|
| **Discipline** | **Women** | **Men** | **Women** | **Men** | **MFR** | ***Z*-statistic** | ***p*-value** |
| AGRI | 11.07 | 13.67 | 10.44 | 15.58 | 1.24 | −7.31 | <0.001 |
| BIO | 8.53 | 11.40 | 9.33 | 13.05 | 1.30 | −12.67 | <0.001 |
| BUS | 7.48 | 7.40 | 7.59 | 6.90 | 1.12 | −1.94 | 0.052 |
| CHEM | 14.19 | 15.42 | 17.40 | 20.12 | 1.17 | −3.30 | <0.001 |
| COMP | 9.36 | 10.28 | 12.85 | 14.00 | 1.17 | −1.28 | 0.201 |
| EARTH | 11.96 | 13.83 | 12.55 | 17.55 | 1.18 | −3.69 | <0.001 |
| ECON | 6.02 | 6.53 | 6.52 | 7.60 | 1.18 | −2.45 | <0.05 |
| ENG | 11.75 | 11.83 | 14.19 | 16.32 | 1.03 | −0.49 | 0.624 |
| ENVIR | 11.29 | 14.23 | 10.87 | 15.33 | 1.25 | −4.18 | <0.001 |
| IMMU | 10.02 | 13.21 | 9.55 | 14.97 | 1.35 | −2.97 | <0.01 |
| MATER | 13.88 | 17.06 | 14.16 | 22.37 | 1.22 | −2.76 | <0.01 |
| MATH | 7.27 | 9.05 | 6.19 | 10.07 | 1.33 | −5.10 | <0.001 |
| MED | 14.48 | 17.34 | 18.47 | 23.48 | 1.24 | −14.00 | <0.001 |
| NEURO | 10.54 | 12.39 | 11.55 | 14.35 | 1.15 | −2.43 | <0.05 |
| PHYS | 23.79 | 23.27 | 48.09 | 52.75 | 0.98 | 0.23 | 0.818 |
| PSYCH | 12.97 | 13.73 | 13.41 | 15.18 | 1.21 | −2.99 | <0.01 |
| TOTAL SOC | 10.12 | 8.97 | 11.36 | 10.80 | 1.05 | −1.16 | 0.246 |
| TOTAL STEMM | 12.97 | 15.46 | 18.13 | 24.87 | 1.24 | −16.40 | <0.001 |

Consequently, the rate can have a value in the 0–100 range, with 0 meaning no publications with international coauthors and 100 meaning all collaborative publications with international coauthors (Table 4).

Our data show that generally in STEMM disciplines, men and women collaborate internationally with very similar intensity, the difference reaching merely 3% (MFR for all STEMM disciplines combined 1.03, $p < 0.001$). For women, the international collaboration rate was about 32, and for men it was about 33. For the vast majority of STEMM disciplines, the difference between men and women was not statistically significant. For the two large disciplines of ENG and MED, women had about 10% higher international collaboration intensity than men (MFR = 0.88, $p < 0.05$ and MFR = 0.92, $p < 0.001$, respectively). However, in two of the three social science disciplines, there were substantial and statistically significant differences between men and women in international collaboration: for BUS, the difference reached 31% in favor of men and for PSYCH, it reached 22% in favor of men. These gender differences are best seen in kernel density plots for PSYCH, BUS, IMMU, and all social sciences combined (TOTAL SOCIAL, Figure 2). The most internationalized disciplines were PHYS and EARTH, with both men and women scientists having about half of their collaborative publications published in international collaboration.

**Table 4.** Average international collaboration rate, all publications for 2000–2023, by discipline and gender

| | Average | | Std | | | | |
|---|---|---|---|---|---|---|---|
| **Discipline** | **Women** | **Men** | **Women** | **Men** | **MFR** | ***Z*-statistic** | ***p*-value** |
| AGRI | 36.31 | 36.90 | 23.93 | 24.94 | 1.02 | −0.64 | 0.522 |
| BIO | 38.31 | 38.20 | 20.69 | 21.19 | 1.00 | 0.20 | 0.841 |
| BUS | 28.64 | 37.44 | 20.38 | 24.56 | 1.31 | −4.49 | <0.001 |
| CHEM | 32.02 | 30.84 | 20.54 | 21.42 | 0.96 | 1.10 | 0.271 |
| COMP | 34.11 | 34.27 | 23.19 | 23.21 | 1.00 | −0.06 | 0.952 |
| EARTH | 48.83 | 49.24 | 23.75 | 24.58 | 1.01 | −0.29 | 0.772 |
| ECON | 37.55 | 38.95 | 26.17 | 25.95 | 1.04 | −0.53 | 0.596 |
| ENG | 28.56 | 25.03 | 22.02 | 20.75 | 0.88 | 2.27 | <0.05 |
| ENVIR | 33.56 | 35.22 | 22.93 | 22.59 | 1.05 | −1.04 | 0.298 |
| IMMU | 34.69 | 37.00 | 18.67 | 21.60 | 1.07 | −1.11 | 0.267 |
| MATER | 31.33 | 31.15 | 20.00 | 22.67 | 0.99 | 0.10 | 0.920 |
| MATH | 42.21 | 45.31 | 27.81 | 27.40 | 1.07 | −1.25 | 0.211 |
| MED | 26.25 | 24.27 | 21.44 | 20.93 | 0.92 | 5.98 | <0.001 |
| NEURO | 34.58 | 35.02 | 19.64 | 19.84 | 1.01 | −0.31 | 0.757 |
| PHYS | 55.52 | 50.52 | 29.48 | 29.11 | 0.91 | 3.45 | <0.001 |
| PSYCH | 23.49 | 28.61 | 21.21 | 22.10 | 1.22 | −3.18 | <0.01 |
| TOTAL SOCIAL | 27.60 | 35.37 | 22.60 | 24.81 | 1.28 | −6.89 | <0.001 |
| TOTAL STEMM | 32.02 | 32.89 | 23.20 | 24.48 | 1.03 | −3.46 | <0.001 |

### 4.3. Publishing Patterns by Journal Type

We wanted to determine whether there was a difference in publishing patterns by journal type (Scopus journal percentile rank) between men and women. Success in publishing means a combination of quantity and quality. This section compares lifetime publication output viewed from the perspective of average Scopus journal percentile rank for men and women within disciplines. We used current journal percentile ranks (2023) retrospectively, assuming that past locations of journals in the Scopus journal architecture were generally similar to their current locations (historical data from Scopus are not available). Scopus locates all its indexed journals in percentiles in the 0–99 range, with most prestigious journals generally located in the 90–99th percentiles (the upper 10% out of about 40,000) or in quartile 1 (the upper 25%).

Our data (Tables 5 and 6) show that the men and women on average published in journals with the same percentile rank, with women publishing in journals with slightly higher percentile ranks in four disciplines, as well as in all social science disciplines combined (MFR = 0.93, $p < 0.001$) and in all STEMM disciplines combined (MFR = 0.98, $p < 0.001$). The most substantial statistically significant difference was for ECON, with women publishing on average in journals with 10% higher journal percentile ranks. The only discipline with slightly higher average percentile ranks for men was BIO (MFR = 1.03, $p < 0.001$). The average journal percentile rank for all STEMM disciplines combined from 2000 to 2023 was 47.96 for women and 46.86 for men, that is, in the middle of the 0–99 range. These results from the accumulation of publications lifetime can be compared with the results for the last five years studied (2019–2023), however.

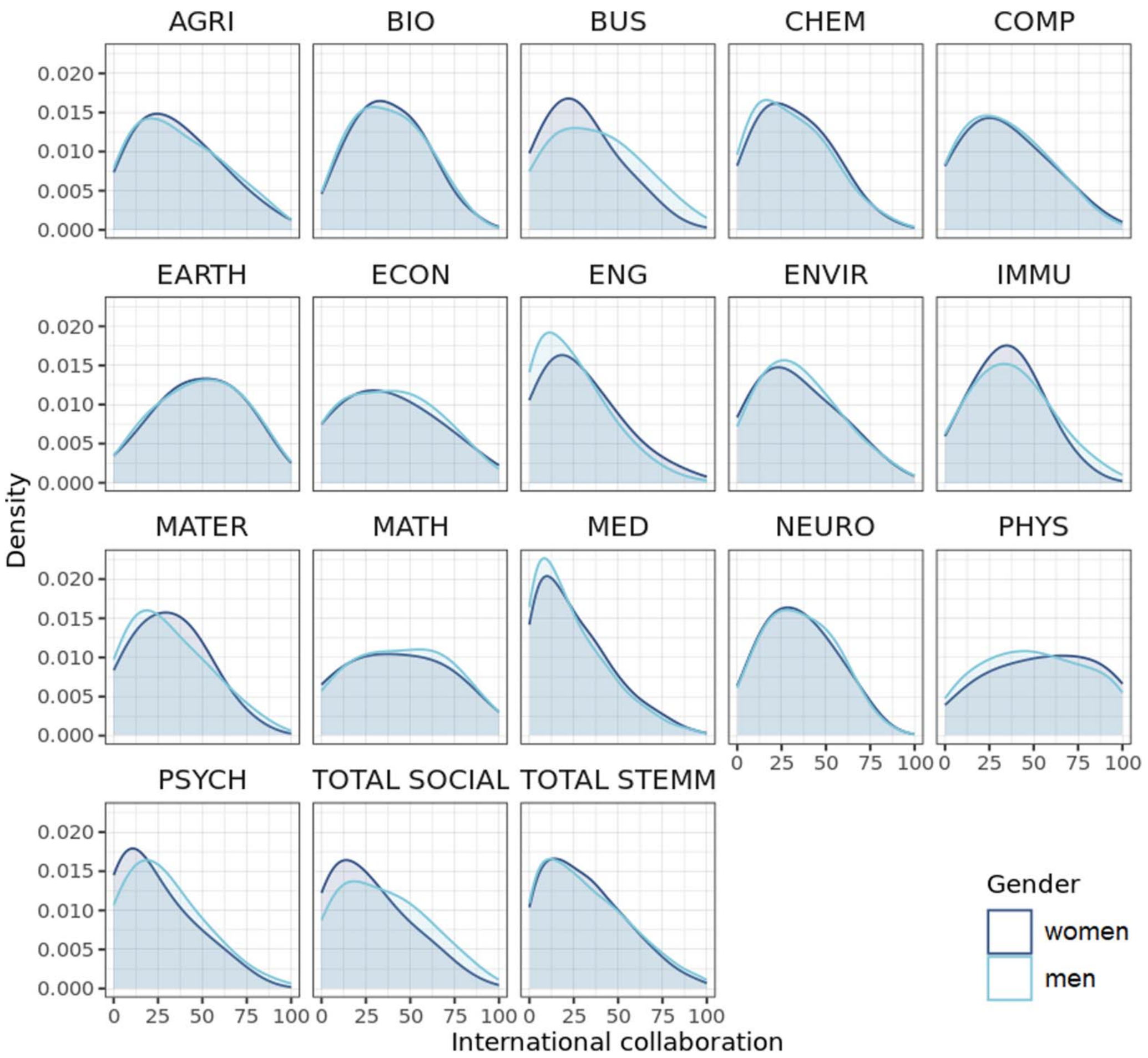

**Figure 2.** Distribution of international collaboration rate, all publications for 2000–2023; kernel density plot by discipline and gender.

Generally, for the five years studied, the same patterns hold in disciplines and in all STEMM disciplines combined: Women on average published in journals that were located slightly higher in Scopus, and wherever the difference between men and women was statistically significant, women on average published in journals with higher Scopus percentile ranks (except BUS).

However, what was interesting was that both men and women published in journals with much higher Scopus percentile ranks compared with the lifetime data: The average was 75 for women and 74 for men. The massive change toward better located journals can best be seen in kernel density plots (Figure 3 vs. Figure 4), with many disciplines reaching the Scopus journal percentile rank of 80. For all disciplines, kernel density plots become left skewed. The change in the distribution of the average Scopus journal percentile rank is impressive and concerns all scientists, regardless of gender. The data clearly show how the scientists, with the passage of time, gradually published in journals located higher in Scopus journal rankings.

**Table 5.** Average journal percentile rank (Scopus), all publications for 2000–2023, by discipline and gender

| | Average | | Std | | | | |
|---|---|---|---|---|---|---|---|
| **Discipline** | **Women** | **Men** | **Women** | **Men** | **MFR** | ***Z*-statistic** | ***p*-value** |
| AGRI | 48.69 | 47.11 | 14.99 | 15.91 | 0.97 | 2.70 | <0.01 |
| BIO | 48.39 | 49.86 | 14.90 | 15.21 | 1.03 | −3.73 | <0.001 |
| BUS | 47.02 | 46.41 | 16.14 | 14.11 | 0.99 | 0.44 | 0.660 |
| CHEM | 47.82 | 47.70 | 15.51 | 16.81 | 1.00 | 0.15 | 0.881 |
| COMP | 49.69 | 48.93 | 16.39 | 17.04 | 0.98 | 0.39 | 0.697 |
| EARTH | 51.33 | 51.60 | 15.70 | 15.98 | 1.01 | −0.29 | 0.772 |
| ECON | 43.94 | 39.62 | 12.82 | 13.51 | 0.90 | 3.32 | <0.001 |
| ENG | 49.55 | 46.23 | 16.51 | 17.60 | 0.93 | 2.80 | <0.01 |
| ENVIR | 53.77 | 53.37 | 14.37 | 15.78 | 0.99 | 0.39 | 0.697 |
| IMMU | 48.13 | 48.39 | 15.03 | 14.07 | 1.01 | −0.17 | 0.865 |
| MATER | 48.56 | 48.17 | 15.35 | 16.94 | 0.99 | 0.29 | 0.772 |
| MATH | 36.03 | 37.30 | 13.03 | 12.64 | 1.04 | −1.09 | 0.276 |
| MED | 47.40 | 44.98 | 16.86 | 17.29 | 0.95 | 9.10 | <0.001 |
| NEURO | 49.10 | 50.69 | 15.26 | 14.60 | 1.03 | −1.45 | 0.147 |
| PHYS | 47.69 | 45.78 | 18.22 | 17.88 | 0.96 | 2.14 | <0.05 |
| PSYCH | 49.31 | 48.87 | 15.15 | 14.97 | 0.99 | 0.39 | 0.697 |
| TOTAL SOCIAL | 47.65 | 44.46 | 15.15 | 14.70 | 0.93 | 4.41 | <0.001 |
| TOTAL STEMM | 47.96 | 46.86 | 16.26 | 16.90 | 0.98 | 6.28 | <0.001 |

Figure 4 shows the steepest kernel density plot for BIO (where the average Scopus journal percentile rank is highest, reaching 80) and the flattest for MATH (where it is 65), in both cases with no statistically significant gender differences.

Publishing in journals located higher, in all the STEMM and social science disciplines studied, clearly came with age and academic experience, and the pattern was obvious for both men and women. Men did not find homes in prestigious journals (e.g., the upper 25%) more quickly: It was only in the last five years that the average Scopus journal percentile rank was in the 75–99 range. Although outliers in publishing patterns are possible, the patterns are generally similar in all disciplines, and publishing in top journals clearly comes with age—which testifies to the cumulative nature of science. As they grow older, survivors publish in increasingly higher prestige journals over their life course—among both men and women.

### 4.4. Publication Productivity

Above, we examined lifetime publishing output (i.e., publication numbers) regardless of journal, following an assumption that "all (Scopus) journals are equal." In this section, we consider journal types as classified by Scopus. To calculate publishing productivity, we used a full counting, journal prestige-normalized approach. Our survivors in science are not in the age in which publication productivity owes to periods when scientists may be expected to experience potential demands of bearing and caring for children. Therefore, survivors are a distinctive group of scientists.

**Table 6.** Average journal percentile rank (Scopus), all publications for 2019–2023, by discipline and gender

| | Average | | Std | | | | |
|---|---|---|---|---|---|---|---|
| **Discipline** | **Women** | **Men** | **Women** | **Men** | **MFR** | ***Z*-statistic** | ***p*-value** |
| AGRI | 75.71 | 72.95 | 14.42 | 16.20 | 0.96 | 4.80 | <0.001 |
| BIO | 80.64 | 80.33 | 12.41 | 12.48 | 1.00 | 0.95 | 0.342 |
| BUS | 77.05 | 80.09 | 17.83 | 14.15 | 1.04 | −2.03 | <0.05 |
| CHEM | 77.05 | 76.87 | 13.77 | 14.31 | 1.00 | 0.25 | 0.806 |
| COMP | 75.09 | 75.03 | 15.23 | 16.03 | 1.00 | 0.03 | 0.976 |
| EARTH | 79.39 | 80.11 | 13.22 | 11.92 | 1.01 | −0.95 | 0.342 |
| ECON | 75.69 | 70.52 | 14.10 | 17.20 | 0.93 | 3.48 | <0.001 |
| ENG | 75.60 | 72.07 | 14.93 | 18.08 | 0.95 | 3.24 | <0.01 |
| ENVIR | 80.91 | 78.44 | 10.88 | 14.59 | 0.97 | 2.93 | <0.01 |
| IMMU | 78.67 | 77.20 | 12.75 | 12.81 | 0.98 | 1.11 | 0.267 |
| MATER | 74.46 | 73.65 | 13.73 | 16.29 | 0.99 | 0.67 | 0.503 |
| MATH | 64.61 | 64.56 | 15.27 | 16.07 | 1.00 | 0.04 | 0.968 |
| MED | 72.83 | 70.04 | 15.17 | 16.32 | 0.96 | 11.42 | <0.001 |
| NEURO | 79.43 | 79.29 | 11.01 | 11.95 | 1.00 | 0.17 | 0.865 |
| PHYS | 78.38 | 77.16 | 13.26 | 13.49 | 0.98 | 1.86 | 0.063 |
| PSYCH | 74.19 | 74.88 | 14.50 | 15.94 | 1.01 | −0.61 | 0.542 |
| TOTAL SOC | 75.27 | 74.70 | 15.46 | 16.43 | 0.99 | 0.75 | 0.453 |
| TOTAL STEMM | 75.41 | 73.96 | 14.66 | 15.83 | 0.98 | 9.05 | <0.001 |

In a full counting approach, as opposed to a fractional counting approach, full credit for a publication goes to all the coauthors. In a journal prestige-normalized approach (as in Kwiek & Roszka, 2024; Kwiek & Szymula, 2025b), as opposed to a prestige nonnormalized approach, the value of a publication depends on the journal location in Scopus journal percentile ranks, which are in the 0–99 percentile range. A solo-authored paper published in a journal located in the 90th percentile rank is valued as 0.9, and a solo-authored paper published in a journal located in the 50th percentile rank is valued as 0.5. If the papers have two coauthors, they are also valued as 0.9 and 0.5, respectively.

Journal-prestige normalization reflects the difference in scientific effort between publishing in lower versus higher prestige journals. Prestige percentile ranks in Scopus are based on citations received in the past four years and reflect the highly stratified nature of academic journals indexed in Scopus. Publishing productivity is calculated as the number of publications (full counting, journal prestige–normalized) divided by the number of years (24 for lifetime productivity, five years for recent productivity in 2019–2023).

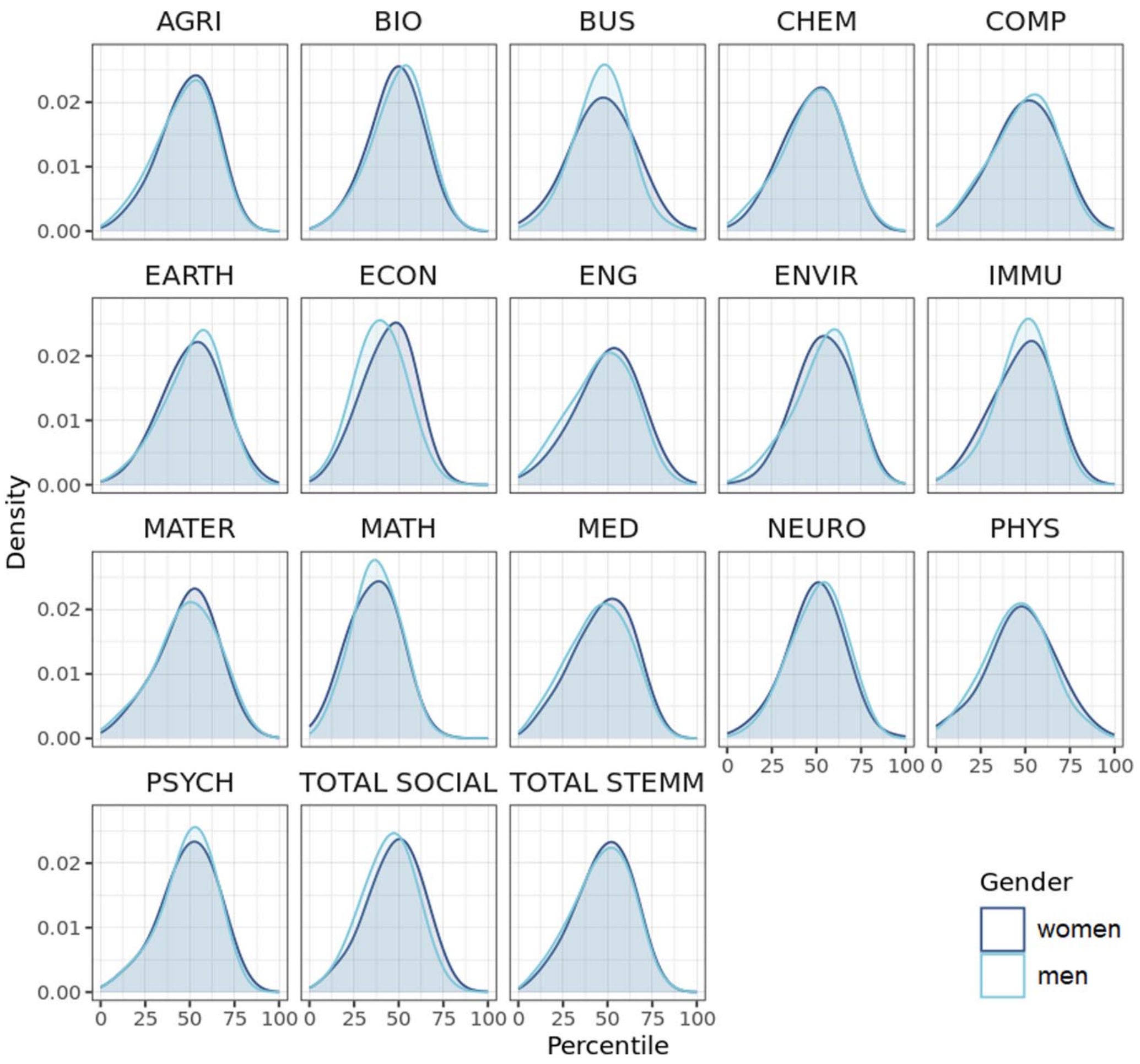

**Figure 3.** Distribution of average journal percentile rank (Scopus), all publications for 2000–2023; kernel density plot by discipline and gender.

The patterns of differences in publishing productivity between men and women from a lifetime perspective and in the recent five years are the same (Tables 7 and 8). Men in STEMM disciplines combined emerged as 23% and 19% more productive, respectively. From a lifetime perspective, men were over 30% more productive than women in three disciplines (MATH MFR = 1.39, $p < 0.001$; BIO MFR = 1.35, $p < 0.001$ and IMMU MFR = 1.33, $p < 0.05$). In the social sciences, a statistically significant difference was observed only in PSYCH, with men on average 21% more productive than women (MFR = 1.21, $p < 0.05$). In all statistically significant cases, men were more productive than women. However, gender differences were slightly smaller in 2019–2023: For all STEMM disciplines combined, the average difference decreased from 23% to 19%, and for all social sciences combined, women were 11% more productive (MFR = 0.89, $p < 0.05$). For the three disciplines with the highest difference in productivity in favor of men over 24 years, the differences were slightly lower (MATH MFR = 1.26, $p < 0.01$; BIO MFR = 1.34, $p < 0.001$ and IMMU MFR = 1.29, $p < 0.05$). The shapes of the distribution of productivity in disciplines in both periods are similar (Figure 5 and Figure 6).

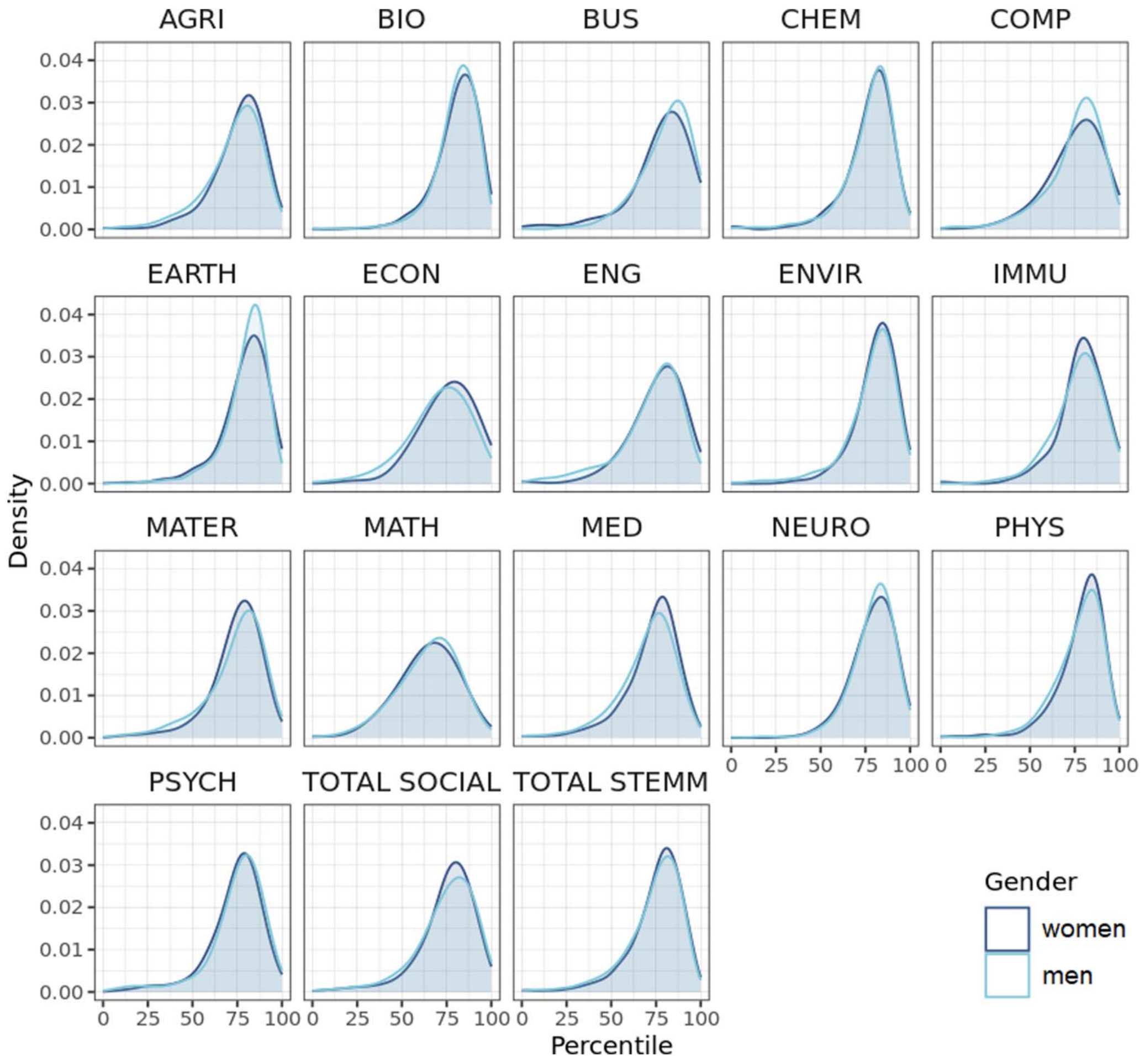

**Figure 4.** Distribution of average journal percentile rank (Scopus), all publications for 2019–2023; kernel density plot by discipline and gender.

However, the values are on average twice as high. Survivors in science, both men and women, were on average twice as productive in the recent period compared with their total productivity. On average, their recent productivity in MATH increased from 0.46 (women) and 0.64 (men) to 0.98 (women) and 1.23 (men), and in the large discipline of BIO, their average productivity increased from 0.74 (women) and 1.00 (men) to 1.40 (women) and 1.87 (men). This substantial increase in publishing productivity is not seen in the shape of plots but in the range, which increased from 0–3 for lifetime productivity to 0–6 for recent productivity (Figure 6).

### 4.5. Being Cited, or the Impact on Global Science

The literature often emphasizes the gender citation gap: Articles authored by women are generally less cited than those authored by men. We calculated the average impact of the articles in terms of citations as viewed through FWCI 4y. The impact is normalized to our disciplines, as in some disciplines, there are on average more citations (e.g., in PHYS), and in others there are on average fewer (e.g., in MATH). Generally, there was no gender difference in impact for all STEMM or all social disciplines combined. Only five disciplines exhibited a statistically significant difference between men and women (BIO, CHEM, EARTH, ENG, and MATH). The largest difference was in MATH with a 20% difference in favor of men (0.89 vs. 1.07, MFR = 1.20, $p < 0.001$; Table 9).

**Table 7.** Average annual publication productivity, all publications for 2000–2023, by discipline and gender

| | Average | | Std | | | | |
|---|---|---|---|---|---|---|---|
| **Discipline** | **Women** | **Men** | **Women** | **Men** | **MFR** | ***Z*-statistic** | ***p*-value** |
| AGRI | 0.82 | 1.02 | 0.73 | 1.07 | 1.24 | −6.08 | <0.001 |
| BIO | 0.74 | 1.00 | 0.71 | 1.04 | 1.35 | −11.62 | <0.001 |
| BUS | 0.54 | 0.59 | 0.45 | 0.49 | 1.09 | −1.20 | 0.230 |
| CHEM | 1.17 | 1.37 | 1.51 | 1.54 | 1.17 | −2.55 | <0.05 |
| COMP | 0.63 | 0.72 | 0.84 | 0.93 | 1.14 | −0.87 | 0.384 |
| EARTH | 0.93 | 1.10 | 0.86 | 1.22 | 1.18 | −3.00 | <0.01 |
| ECON | 0.46 | 0.50 | 0.41 | 0.50 | 1.09 | −0.93 | 0.352 |
| ENG | 0.84 | 0.82 | 0.92 | 1.05 | 0.98 | 0.30 | 0.764 |
| ENVIR | 0.90 | 1.15 | 0.79 | 1.19 | 1.28 | −3.86 | <0.001 |
| IMMU | 0.83 | 1.10 | 0.73 | 1.31 | 1.33 | −2.50 | <0.05 |
| MATER | 1.11 | 1.39 | 1.20 | 1.66 | 1.25 | −2.52 | <0.05 |
| MATH | 0.46 | 0.64 | 0.37 | 0.68 | 1.39 | −4.42 | <0.001 |
| MED | 1.01 | 1.21 | 1.22 | 1.56 | 1.20 | −9.35 | <0.001 |
| NEURO | 0.92 | 1.08 | 0.88 | 1.08 | 1.17 | −2.29 | <0.05 |
| PHYS | 2.39 | 2.30 | 5.75 | 5.92 | 0.96 | 0.32 | 0.749 |
| PSYCH | 0.92 | 1.11 | 0.94 | 1.19 | 1.21 | −2.39 | <0.05 |
| TOTAL SOCIAL | 0.73 | 0.71 | 0.77 | 0.82 | 0.97 | 0.53 | 0.596 |
| TOTAL STEMM | 0.98 | 1.21 | 1.56 | 2.27 | 1.23 | −11.81 | <0.001 |

Our data show similar journal and citation patterns, and the two are clearly linked. When papers written by men and women are published on average in journals with similar locations in Scopus journal ranks (based on citation data), they are cited in a comparable manner. Similar average publication patterns lead to similar average citation patterns. The data for the recent five years are even more convincing: There are no statistically significant differences for any disciplines at all except for MATH, where the average impact of men was 26% higher than the average impact of women (FWCI 4y women 0.70, FWCI 4y men 0.88, MFR = 1.26, $p < 0.01$, Table S1 in the Supplementary material).

### 4.6. Collaboration—Team Formation

Our data also allowed us to compare team formation by men and women survivors, both longitudinally in 2000–2023 and in 5 recent years (2019–2023). Both in social disciplines combined and in STEMM disciplines combined, the difference between average median team size for men and women was statistically significant and favored women. Women also formed on average slightly larger teams in the seven disciplines for which the gender difference was statistically significant. In the social science disciplines combined, women on average were members of teams of 3.20 scientists and men on average 2.83 (MFR = 0.88, $p < 0.001$), and in the STEMM disciplines combined, the teams are on average 6.74 scientists for women and 6.24 for men (MFR = 0.93, $p < 0.001$). Thus, across the board, women worked in slightly larger teams. The pattern was identical for the recent 5 years: women worked in slightly bigger teams in both the social sciences combined and the STEMM disciplines combined, and the MFR was the same, despite team size averages being slightly larger in both cases (3.80 for women and 3.34 for men, and 7.54 for women and 7.01 for men, respectively; Table S2 in ESM).

**Table 8.** Average publication productivity, all publications for 2019–2023, by discipline and gender

| Discipline | Average | | Std | | MFR | Z-statistic | p-value |
|---|---|---|---|---|---|---|---|
| | Women | Men | Women | Men | | | |
| AGRI | 1.73 | 2.10 | 1.70 | 2.53 | 1.21 | −4.79 | <0.001 |
| BIO | 1.40 | 1.87 | 1.54 | 2.21 | 1.34 | −9.80 | <0.001 |
| BUS | 1.21 | 1.21 | 1.25 | 1.14 | 1.00 | 0.00 | 1.000 |
| CHEM | 2.30 | 2.51 | 2.97 | 3.45 | 1.09 | −1.30 | 0.194 |
| COMP | 1.49 | 1.62 | 2.29 | 2.42 | 1.09 | −0.47 | 0.638 |
| EARTH | 1.97 | 2.30 | 2.11 | 3.00 | 1.17 | −2.37 | <0.05 |
| ECON | 0.93 | 0.96 | 1.00 | 1.29 | 1.03 | −0.28 | 0.779 |
| ENG | 1.87 | 1.80 | 2.49 | 2.67 | 0.96 | 0.39 | 0.697 |
| ENVIR | 1.87 | 2.34 | 1.85 | 2.66 | 1.25 | −3.17 | <0.01 |
| IMMU | 1.60 | 2.06 | 1.54 | 2.37 | 1.29 | −2.25 | <0.05 |
| MATER | 2.20 | 2.71 | 2.45 | 3.79 | 1.23 | −2.15 | <0.05 |
| MATH | 0.98 | 1.23 | 0.91 | 1.48 | 1.26 | −2.64 | <0.01 |
| MED | 2.21 | 2.58 | 2.99 | 3.73 | 1.17 | −7.15 | <0.001 |
| NEURO | 1.73 | 2.02 | 1.99 | 2.44 | 1.17 | −1.84 | 0.066 |
| PHYS | 4.04 | 3.96 | 8.86 | 9.74 | 0.98 | 0.18 | 0.857 |
| PSYCH | 1.99 | 2.15 | 2.12 | 2.41 | 1.08 | −0.95 | 0.342 |
| TOTAL SOC | 1.57 | 1.39 | 1.80 | 1.75 | 0.89 | 2.10 | <0.05 |
| TOTAL STEMM | 2.03 | 2.42 | 3.06 | 4.29 | 1.19 | −10.41 | <0.001 |

### 4.7. Publishing Intensity—Median Publishing Breaks

Our data set permitted us to compare the publishing intensity of men and women: What percentage published at least one article annually for 21–24 consecutive years? Zero publishing breaks means at least one publication every year; one publishing break means one year in2000–2023 without a publication; and having two publishing breaks means two years without a publication. Our results (Table 11) show that about 80% of the scientists in STEMM (79.55% of women and 83.28% of men, $p < 0.001$) and about 70% of scientists in the social sciences (72.04% of women and 73.11% of men, no statistical significance) published continuously (these results fill a knowledge gap mentioned by Ioannidis et al., 2014, p. 1: "there is no data on what proportion of scientists manages to publish each and every year over long periods of time"). This represents powerful, uninterrupted occupation with research for a very long time.

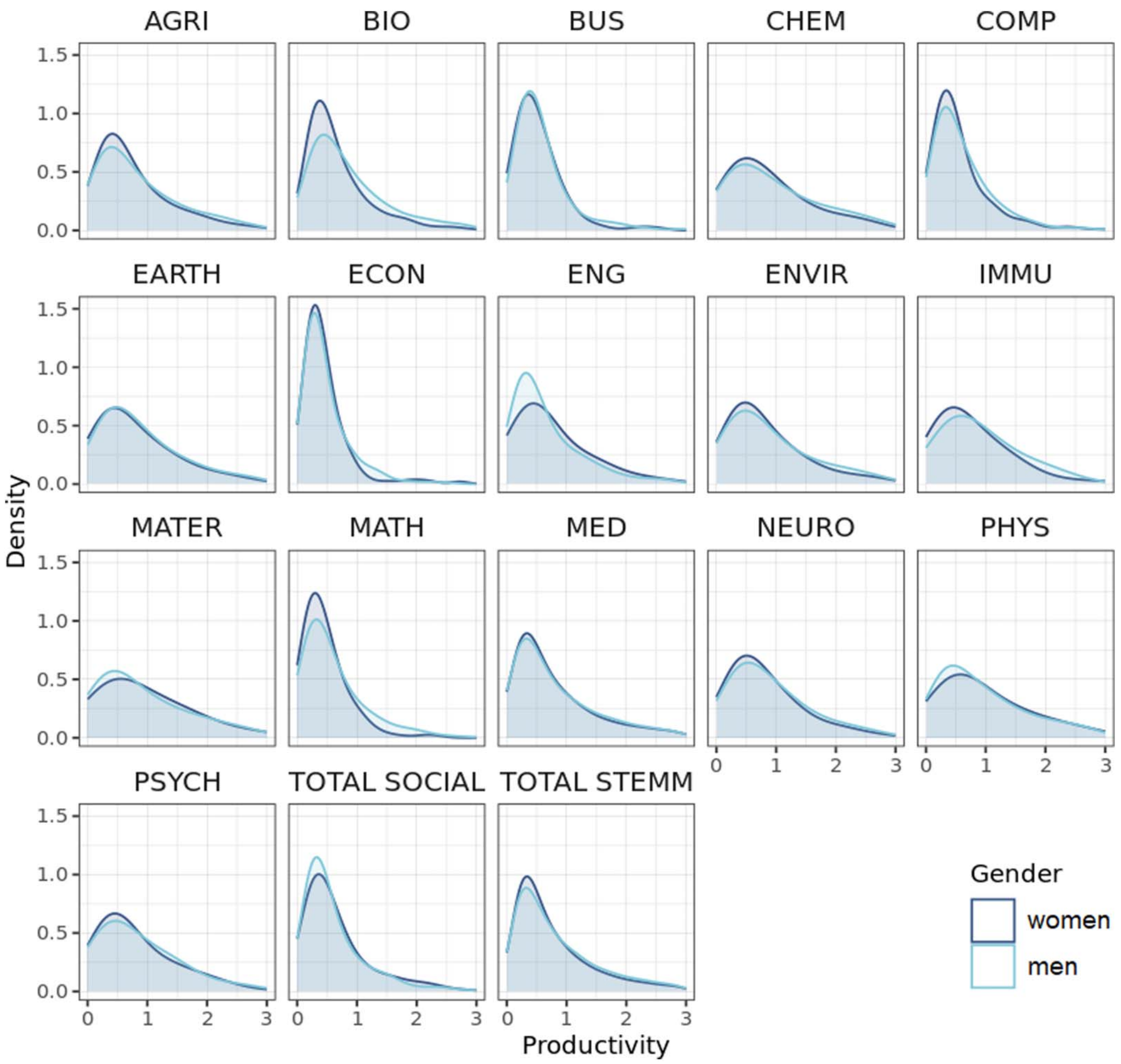

**Figure 5.** Distribution of (annual) publication productivity, all publications for 2000–2023; kernel density plot by discipline and gender.

The gender difference in median publishing breaks was statistically significant for STEMM disciplines but not for the social sciences. The highest publishing intensity was observed for PHYS, in which about 90% of the men and women published continuously for 21–24 years, followed by CHEM with about 85% for women and 88% for men, and IMMU with about 89% for men. In only five disciplines was the gender difference in percentages of zero publishing breaks statistically significant, including in the two largest, MED and BIO. In each case, the percentage was higher for men. The lowest percentage was in two social sciences, BUS and ECON. Overall, the survivors in science published continuously—which may be one reason for their academic success (job stability). No matter what happened in their personal lives, they delivered publications almost every year, their productivity increased—and they published in ever more prestigious journals.

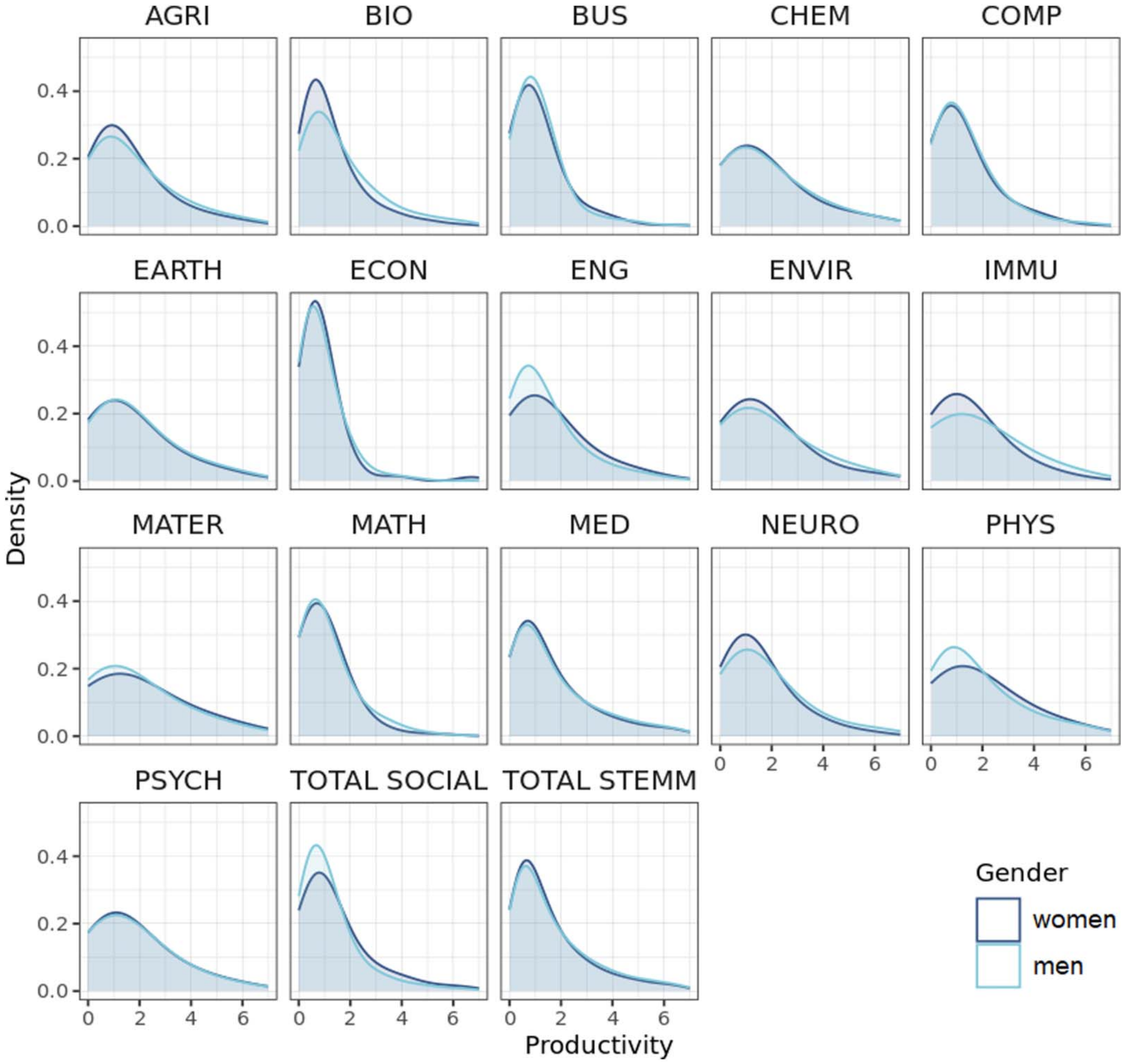

**Figure 6.** Distribution of publication productivity, all publications for 2019–2023; kernel density plot by discipline and gender.

### 4.8. A Cross-Disciplinary Summary

Table 12 summarizes the male-to-female ratios (M/F) across all disciplines and all dimensions of gender disparities in science considered in this study, together with *p*-values from statistical tests comparing men and women. Because the ratios are defined as M/F, values above 1 indicate higher average values for men, while values below 1 indicate higher average values for women.

The largest and statistically significant gender differences in lifetime scholarly output were observed in IMMU, MATH, and BIO (approximately 30% in favor of men). The largest difference in international collaboration was found in BUS (31% in favor of men). In journal prestige (Scopus journal percentile rank), a notable difference was observed only in ECON (10% in favor of women). In citation impact (FWCI 4y), the strongest disparities occurred in MATH (20% in favor of men). Women had larger average team sizes in both the combined social sciences (12% in favor of women) and in the combined STEMM sciences (7% in favor of women). This finding runs somehow counter to the conventional wisdom of team formation in science in general but is in line with what Fox and Mohapatra (2007) found using a national mail survey of 1,215 faculty in US doctoral-granting departments of computer science, chemistry, electrical engineering, microbiology, and physics. Seeking correlations between team composition and productivity, they found that the key variable is the interaction between being a male faculty and having higher numbers of male graduate students on the team. This interaction remains significant in its relationship to productivity with the addition of other variables (collaboration, work practice, work climate, and context; Fox and Mohapatra, 2007). The highest gender disparities in annual productivity appeared in MATH (39% in favor of men).

**Table 9.** Average FWCI 4y, all publications for 2000–2023, by discipline and gender

| | Average | | Std | | | | |
|---|---|---|---|---|---|---|---|
| Discipline | Women | Men | Women | Men | MFR | *Z*-statistic | *p*-value |
| AGRI | 1.24 | 1.22 | 0.81 | 0.78 | 0.98 | 0.65 | 0.516 |
| BIO | 1.83 | 2.01 | 1.87 | 3.26 | 1.10 | −2.74 | <0.01 |
| BUS | 1.76 | 1.75 | 1.21 | 1.41 | 0.99 | 0.09 | 0.928 |
| CHEM | 1.30 | 1.39 | 0.84 | 0.82 | 1.07 | −2.09 | <0.05 |
| COMP | 1.53 | 1.62 | 1.21 | 2.13 | 1.06 | −0.41 | 0.682 |
| EARTH | 1.53 | 1.66 | 0.83 | 1.22 | 1.08 | −2.34 | <0.05 |
| ECON | 1.51 | 1.30 | 1.29 | 1.13 | 0.86 | 1.66 | 0.097 |
| ENG | 1.41 | 1.25 | 0.93 | 0.85 | 0.89 | 2.44 | <0.05 |
| ENVIR | 1.52 | 1.43 | 0.97 | 0.82 | 0.94 | 1.39 | 0.165 |
| IMMU | 1.63 | 1.56 | 1.27 | 1.03 | 0.96 | 0.58 | 0.562 |
| MATER | 1.20 | 1.27 | 0.90 | 1.03 | 1.06 | −0.89 | 0.373 |
| MATH | 0.89 | 1.07 | 0.48 | 0.68 | 1.20 | −3.79 | <0.001 |
| MED | 1.80 | 1.77 | 2.45 | 3.17 | 0.98 | 0.69 | 0.490 |
| NEURO | 1.50 | 1.67 | 0.88 | 2.70 | 1.11 | −1.31 | 0.190 |
| PHYS | 2.10 | 2.05 | 2.23 | 2.01 | 0.98 | 0.46 | 0.646 |
| PSYCH | 1.43 | 1.44 | 0.99 | 0.93 | 1.01 | −0.14 | 0.889 |
| TOTAL SOCIAL | 1.54 | 1.47 | 1.12 | 1.18 | 0.95 | 1.27 | 0.204 |
| TOTAL STEMM | 1.70 | 1.69 | 2.05 | 2.53 | 0.99 | 0.42 | 0.674 |

The most pronounced and consistently significant differences across disciplines concerned lifetime scholarly output (nonnormalized publication counts) and annual productivity (full counting, journal prestige-normalized). Both are indicators closely related to academic promotion and retention (Diezmann & Grieshaber, 2019; Spoon et al., 2023). In BIO, IMMU, and MATH, men's lifetime output exceeded women's by about 30%, and by about 20% in AGRI, MATER, MED, and PSYCH. Annual publication productivity showed an almost identical pattern: approximately 30% higher for men in BIO, MATH, and IMMU, and about 20% higher in AGRI, ENVIR, MATER, MED, and PSYCH. For all STEMM disciplines combined, the gender differences were about one quarter for both lifetime scholarly output (M/F = 1.24) and annual productivity (M/F = 1.23), all statistically significant at $p < 0.001$.

**Table 10.** Average team size, all publications for 2000–2023, by discipline and gender Q3

| | Average | | Std | | | | |
|---|---|---|---|---|---|---|---|
| **Discipline** | **Women** | **Men** | **Women** | **Men** | **MFR** | ***Z*-statistic** | ***p*-value** |
| AGRI | 5.21 | 4.98 | 1.62 | 1.53 | 0.96 | 3.78 | <0.001 |
| BIO | 7.50 | 7.33 | 1.93 | 1.95 | 0.98 | 3.34 | <0.001 |
| BUS | 2.81 | 2.77 | 0.66 | 0.70 | 0.99 | 0.66 | 0.509 |
| CHEM | 5.49 | 5.32 | 1.50 | 1.43 | 0.97 | 2.22 | <0.05 |
| COMP | 3.85 | 3.76 | 1.17 | 1.12 | 0.98 | 0.69 | 0.490 |
| EARTH | 5.52 | 5.54 | 1.93 | 1.99 | 1.00 | −0.17 | 0.865 |
| ECON | 2.40 | 2.39 | 0.63 | 0.62 | 1.00 | 0.16 | 0.873 |
| ENG | 3.94 | 4.03 | 1.22 | 1.34 | 1.02 | −1.02 | 0.308 |
| ENVIR | 5.09 | 4.90 | 1.55 | 1.55 | 0.96 | 1.76 | 0.078 |
| IMMU | 7.01 | 7.01 | 1.85 | 1.84 | 1.00 | 0.00 | 1.000 |
| MATER | 5.18 | 5.33 | 1.42 | 1.53 | 1.03 | −1.23 | 0.219 |
| MATH | 2.53 | 2.39 | 0.70 | 0.79 | 0.94 | 2.17 | <0.05 |
| MED | 7.27 | 7.28 | 2.20 | 2.11 | 1.00 | −0.3 | 0.764 |
| NEURO | 6.12 | 5.75 | 1.94 | 1.95 | 0.94 | 2.62 | <0.01 |
| PHYS | 6.89 | 6.47 | 2.59 | 2.62 | 0.94 | 3.29 | <0.01 |
| PSYCH | 3.69 | 3.49 | 1.28 | 1.14 | 0.95 | 2.21 | <0.05 |
| TOTAL SOCIAL | 3.20 | 2.83 | 1.17 | 0.95 | 0.88 | 6.99 | <0.001 |
| TOTAL STEMM | 6.74 | 6.24 | 2.29 | 2.35 | 0.93 | 20.35 | <0.001 |

Importantly, the analysis for the most recent period (2019–2023) showed gender differences very similar in magnitude to those for the entire study window. This suggests that the gender productivity gap does not originate solely from early-career differences (e.g., family or child-related interruptions among women) and does not diminish over time, contrary to what might be expected. The hypothesis that the productivity gap is primarily rooted in early career stages and later disappears (Morgan et al., 2021; Tatarini, Gorodetskaya, & Vitali, 2024) was not supported. For all STEMM disciplines combined, the M/F ratios for lifetime scholarly output for all years and for 2019–2023 were identical (1.24), and the corresponding ratios for annual publication productivity were 1.23 and 1.19, respectively (all statistically significant at $p < 0.001$).

**Table 11.** Median publishing breaks, all publications for 2000–2023, by discipline and gender (in %)

| | Women | | | | | Men | | | | | Gender difference | |
|---|---|---|---|---|---|---|---|---|---|---|---|---|
| **Discipline** | **0** | **1** | **2** | **3** | **4+** | **0** | **1** | **2** | **3** | **4+** | $\chi^2$ | ***p*-value** |
| AGRI | 79.00 | 17.82 | 2.41 | 0.67 | 0.10 | 81.94 | 15.07 | 2.74 | 0.25 | 0.00 | 1,073,492.0 | 0.054 |
| BIO | 75.69 | 21.52 | 2.49 | 0.25 | 0.05 | 84.16 | 13.80 | 1.89 | 0.11 | 0.04 | 4,822,968.0 | <0.001 |
| BUS | 67.91 | 28.88 | 2.67 | 0.53 | 0.01 | 75.94 | 18.72 | 5.08 | 0.26 | 0.00 | 37,451.0 | 0.076 |
| CHEM | 85.05 | 12.52 | 2.43 | 0.00 | 0.00 | 88.97 | 9.34 | 1.61 | 0.08 | 0.00 | 345,311.5 | <0.05 |
| COMP | 70.45 | 24.29 | 4.45 | 0.40 | 0.41 | 75.87 | 19.91 | 3.58 | 0.55 | 0.09 | 141,953.0 | 0.076 |
| EARTH | 78.39 | 17.45 | 3.13 | 0.78 | 0.25 | 83.85 | 13.28 | 2.52 | 0.26 | 0.09 | 233,432.5 | <0.05 |
| ECON | 59.68 | 29.84 | 8.87 | 1.61 | 0.00 | 66.26 | 28.48 | 4.65 | 0.61 | 0.00 | 33,153.0 | 0.099 |
| ENG | 78.83 | 18.02 | 2.70 | 0.00 | 0.45 | 77.78 | 18.24 | 3.31 | 0.67 | 0.00 | 195,440.0 | 0.687 |
| ENVIR | 75.68 | 21.28 | 2.70 | 0.34 | 0.00 | 79.33 | 16.72 | 3.08 | 0.59 | 0.28 | 104,268.0 | 0.253 |
| IMMU | 79.43 | 17.14 | 2.86 | 0.57 | 0.00 | 88.94 | 11.06 | 0.00 | 0.00 | 0.00 | 19,135.5 | <0.01 |
| MATER | 85.16 | 12.64 | 2.20 | 0.00 | 0.00 | 87.54 | 10.36 | 2.10 | 0.00 | 0.00 | 57,552.5 | 0.411 |
| MATH | 74.05 | 20.89 | 5.06 | 0.00 | 0.00 | 77.05 | 20.03 | 2.74 | 0.17 | 0.01 | 47,705.5 | 0.374 |
| MED | 80.17 | 16.32 | 2.98 | 0.40 | 0.13 | 83.09 | 14.10 | 2.42 | 0.32 | 0.07 | 35,840,814.0 | <0.001 |
| NEURO | 82.94 | 14.72 | 2.01 | 0.33 | 0.00 | 83.20 | 15.04 | 1.76 | 0.00 | 0.00 | 76,819.0 | 0.896 |
| PHYS | 90.26 | 8.11 | 1.22 | 0.20 | 0.21 | 88.84 | 9.82 | 1.03 | 0.20 | 0.11 | 616,547.5 | 0.372 |
| PSYCH | 78.67 | 17.87 | 3.17 | 0.29 | 0.00 | 79.31 | 15.38 | 4.77 | 0.53 | 0.01 | 65,551.5 | 0.943 |
| TOTAL SOCIAL | 72.04 | 23.25 | 4.10 | 0.61 | 0.00 | 73.11 | 21.59 | 4.82 | 0.48 | 0.00 | 413,510.5 | 0.687 |
| TOTAL STEMM | 79.55 | 17.20 | 2.78 | 0.37 | 0.10 | 83.28 | 14.08 | 2.29 | 0.28 | 0.07 | 181,342,036.0 | <0.001 |

Although gender differences in both lifetime output and annual productivity were statistically significant in STEMM as a whole, this pattern did not hold in the social sciences as a whole. In PSYCH, men were 21% more productive and had 21% higher total output, while in ECON they had 18% higher lifetime output but no significant difference in annual productivity. In BUS, no gender differences were observed in either output or productivity.

The *p*-values accompanying the M/F ratios refer to two-sample tests comparing the distributions of the underlying raw indicators for men and women (e.g., publication counts, collaboration rates, journal percentiles, citation impact). Conceptually, these tests assess whether the male-to-female ratio differs from 1, that is, whether there is any statistically significant difference between men and women for a given indicator. The tests were two-sided ($H_1$: M/F $\neq$ 1). Given the extremely low *p*-values in many cases, the conclusions would remain unchanged if one-sided tests were applied.

**Table 12.** Male-to-female ratios (M/F) for selected indicators and *p*-values from tests comparing men and women, 2000–2023

| Discipline | Lifetime scholarly output (M/F) | International collaboration (M/F) | Scopus journal percentile (M/F) | Annual publication productivity (M/F) | FWCI 4y (M/F) | Team size (M/F) |
|---|---|---|---|---|---|---|
| AGRI | 1.24*** | 1.02 | 0.97** | 1.24*** | 0.98 | 0.96*** |
| BIO | 1.30*** | 1.00 | 1.03*** | 1.35*** | 1.10** | 0.98*** |
| BUS | 1.12 | 1.31*** | 0.99 | 1.09 | 0.99 | 0.99 |
| CHEM | 1.17*** | 0.96 | 1.00 | 1.17* | 1.07* | 0.97* |
| COMP | 1.17 | 1.00 | 0.98 | 1.14 | 1.06 | 0.98 |
| EARTH | 1.18*** | 1.01 | 1.01 | 1.18** | 1.08* | 1.00 |
| ECON | 1.18* | 1.04 | 0.90*** | 1.09 | 0.86 | 1.00 |
| ENG | 1.03 | 0.88* | 0.93** | 0.98 | 0.89* | 1.02 |
| ENVIR | 1.25*** | 1.05 | 0.99 | 1.28*** | 0.94 | 0.96 |
| IMMU | 1.35** | 1.07 | 1.01 | 1.33* | 0.96 | 1.00 |
| MATER | 1.22** | 0.99 | 0.99 | 1.25* | 1.06 | 1.03 |
| MATH | 1.33*** | 1.07 | 1.04 | 1.39*** | 1.20*** | 0.94* |
| MED | 1.24*** | 0.92*** | 0.95*** | 1.20*** | 0.98 | 1.00 |
| NEURO | 1.15* | 1.01 | 1.03 | 1.17* | 1.11 | 0.94** |
| PHYS | 0.98 | 0.91*** | 0.96** | 0.96 | 0.98 | 0.94** |
| PSYCH | 1.21** | 1.22** | 0.99 | 1.21* | 1.01 | 0.95* |
| TOTAL SOCIAL | 1.05 | 1.28*** | 0.93*** | 0.97 | 0.95 | 0.88*** |
| TOTAL STEMM | 1.24*** | 1.03*** | 0.98*** | 1.23*** | 0.99 | 0.93*** |

***$p \leq 0.001$, **$p \leq 0.01$, *$p \leq 0.05$

## 5. DISCUSSION AND CONCLUSIONS

We examined gender disparities in publishing intensity (lifetime scholarly output), international collaboration, journal publishing patterns, publishing productivity, citations, team formation, and yearly publishing patterns (or publishing breaks). We had a clearly defined sample: all scientists from 16 disciplines (13 STEMM and three social sciences) who had at least 10 journal articles in their portfolios, started their publishing career in 2000, had at least one publication in 2020–2023, and had an affiliation in an OECD country. We used microdata at the scientist level ($n$ = 41,424) and those on over two million publications and more than 70 million cited references. Every scientist in our sample had an unambiguously ascribed gender, a discipline, an academic age (years of publishing experience), and a country affiliation. We had publications with their metadata, cited references from all publications with their metadata, the citation impact for four years (FWCI 4y) for each publication based on their citation numbers, the number of collaborators in collaborative articles, and the Scopus journal percentile rank for each publication, locating it in a Scopus-based journal prestige system of about 40,000 journals (CiteScore percentile ranks).

Following men and women scientists for up to 24 years (2000–2023 inclusive), using cumulative data for the whole period and the data for the past five years (2019–2023), and working within disciplines, we were able to revisit several traditional gender gaps in science with respect to a highly selective segment of science: "survivors in science."

Conclusions on survivors do not refer to all scientists; they refer only to those successful few (as we have shown elsewhere in detail, there are about 30% of scientists who avoid attrition after 19 years of work across 16 STEM disciplines in 38 OECD countries, with significantly higher probabilities of exit for women in most disciplines; see Kwiek & Szymula, 2025a). Our data on survivors clearly show only a gender publication productivity gap, for both lifetime scholarly output and annual productivity, consistently with wide literature on the general science population (e.g., Abramo et al., 2019; Fox & Mohapatra, 2007; Fox & Nikivincze, 2021; Larivière et al., 2013; Mihaljević-Brandt et al., 2016; Nielsen, 2016; Nygaard, Aksnes, & Piro, 2022; Toutkoushian & Bellas, 1999); no gender international collaboration gap, gender journal selection gap, gender citation gap, or gender team formation gap was observed, despite wide discussions in the literature on the general science population.

Our evidence tends to suggest that these gaps do not exist among a selective segment of "survivors in science"—which does not support the argument that these gaps do not exist in science more broadly. There is no conflict between a large literature on collaboration gap in science (e.g., Abramo et al., 2013; Ackers, 2008; Bear & Wooley, 2011; Elsevier, 2020; Fell & Koenig, 2016; Fox et al., 2017; Uhly et al., 2017) and our findings because our findings apply to an entirely different sample of scientists.

The same sample-related limitation refers to wide literature on gender gaps in citations (Aksnes et al., 2011; Hutson, 2006; King et al., 2017; Maliniak et al., 2013; Mishra et al., 2018; Thelwall, 2020), journal selection (Chaoqun, Basson et al., 2025; Mihaljević-Brandt et al., 2016), and team formation (Feenay & Bernal, 2010; Heffner, 1979), including gender-homophily (Ghiasi et al., 2018; Kegen, 2013; Kwiek & Roszka, 2021b; Lerchenmueller et al., 2019).

This wider literature on various aspects of gender inequalities in science provides an excellent context to our findings, without referring directly to the science segment we studied. Unfortunately, we were unable to capture in this research several other fundamental dimensions of gender inequality in academic work: recognition in team production (Sarsons, 2017; Sarsons et al., 2020; Heffner, 1979), hiring and promotion (Clauset et al., 2015; Suer, 2023; Weisshaar, 2017), academic salaries (Barbezat & Hughes, 2005; Ward & Sloane, 2000) and grant allocation (Cruz-Castro & Sanz-Menendez, 2019). There is a clear dividing line between what can be derived from bibliometric data sources and what requires data from surveys, interviews, grant repositories, and national registries of scientists.

Our study was cross-national (but still limited to 38 OECD countries) and cross-disciplinary, and its cohort and longitudinal nature allowed us to compare "apples with apples" (scientists from the same cohort and discipline over time; see Nygaard et al., 2022), as opposed to cross-sectional studies that include scientists from different cohorts. The scientists in our sample had the same lengths of publishing experience.

The literature shows that women in general are less inclined to engage in international collaboration than men (Ackers, 2008; Aksnes et al., 2019; Fox et al., 2017; Kwiek & Roszka, 2021a; Uhly et al., 2017). Our analysis of survivors does not confirm this. Our data indicate that men and women survivors collaborate internationally with the same intensity, the difference for all STEMM disciplines combined reaching only 3%. For the vast majority of STEMM disciplines, the difference between men and women was not statistically significant. Indeed, for the two large disciplines of ENG and MED, women exhibited 10% higher international collaboration intensity. For some reason, however, in the social sciences, the rates for men were 31% higher in BUS and 22% higher in PSYCH. Our findings' incompatibility with literature on women and international collaboration may be caused by the sample: Our study examined only late-career scientists who persisted in science for a long time. Perhaps women survivors successfully imitated men survivors as much as possible—which proved possible in all dimensions except raw scholarly output and journal prestige-normalized productivity, where gender gaps were observed.

The literature also generally suggests that women tend to publish in lower-ranked journals compared with men, either submitting papers less often to top journals or having them rejected more often by top journals (Mihaljević-Brandt et al., 2016). Again, our data on survivors do not support this. We had a huge data set of more than two million papers closely linked to Scopus journals percentile ranks. It showed that on average, men and women in all the disciplines combined published in journals with the same percentile rank. In fact, women in five disciplines published in journals with slightly higher percentile ranks. This finding is especially surprising in the context of MATH and COMP, in which women are reported to publish in less prestigious journals and papers in conference proceedings (Dabo-Niang, Esteban et al., 2024; Mihaljević-Brandt et al., 2016). In ECON, women on average publish in a 10% higher percentile rank of journals—which is a substantial difference. Scopus journal percentile rankings have limitations (e.g., the volatility of journals with very small numbers of publications per year), but it is the largest citation-driven system that links journals, papers, and citations. Finally, our data clearly show that publishing experience comes with age, and the probability of publishing in top-tier journals increases with (academic) age in all the disciplines studied. Our comparison of the average lifetime percentile rank and the average percentile rank for the last five years indicates the cumulative nature of science and a lack of any substantial gender differences in publishing patterns.

Our findings on survivors also go against traditional literature indicating a gender citation gap, or women being cited less frequently than men (Madison & Fahlman, 2020; Maliniak et al., 2013; Sugimoto & Larivière, 2023; Thelwall, 2020). For survivors, we observed no gender difference in impact on scholarly literature: the average impact in terms of citations, namely the FWCI 4y. For all STEMM disciplines combined and all social science combined, no gender differences exist. The difference was statistically significant in only one third of disciplines, and the largest difference is in MATH, with 20% difference in favor of men. We can assume that similar journal patterns for men and women lead to similar citation patterns. Finally, women survivors form slightly larger research teams. Team formation is extremely important for academic success, and it could be expected that women have more problems with working in larger teams (Bear & Woolley, 2011; Kwiek & Roszka, 2021b). Our data clearly show this is not the case for late-career scientists: There is not a single discipline in which men survivors form larger teams (with the difference statistically significant).

High publishing intensity matters for academic success. Annual productivity shows the number of publications in a year; publishing intensity indicates how many years one publishes uninterruptedly. Our data show that 80% of survivor scientists in STEMM disciplines and about 70% in the social science disciplines, men and women alike, managed to publish continuously in 2000–2023, every year. Importantly, there were no statistically significant differences between men and women. In PHYS, CHEM, and IMMU, 85%–90% of scientists published every year. Our data suggest the importance of publishing every year. Perhaps colleagues of our survivors in science who published less and published less often, published in lower-tier journals, collaborated internationally less, and had smaller teams (and consequently, for all these reasons, were less cited) had to leave academic science (Preston, 2004; White-Lewis et al., 2023; Wohrer, 2014).

The study would be more specific if we had birth cohorts (see Rowland, 2014, pp. 119–178; Wachter, 2014, pp. 30–47) at our disposal; however, to have all scientists of the same biological age in a sample is not possible in cross-national studies (38 countries); instead of biological age, we used its proxy, academic age, or the time passed since the first indexed publication. We have shown elsewhere, using a population of 100,000 scientists, that in STEMM disciplines in Poland, academic age is highly correlated with biological age. A systematic analysis of a national system showed that the correlation between the two types of age is very high, the correlation coefficients being in the 80–90% range (e.g., 0.89 for CHEM, 0.88 for PHYS, 0.85 for MATH, and 0.90 for IMMU; Kwiek & Roszka, 2022a).

Methodologically, our use of disciplines is also based on proxies: Because it is impossible to ascribe scientists from 38 countries to unique national systems of disciplines (e.g., in national research assessment exercises), our approach was to keep the commercial ASJC system used by Scopus in its data set. However, our method was fine-grained: With the average lifetime (24-year) output being about 50 papers per scientist (54 for men and 44 for women; Table 2), we had on average about 1,800 cited references per scientist to determine their discipline. In total, we used more than 70 million cited references ($n$ = 73,118,395) to determine the disciplines of 41,424 scientists. Our simplifying assumption was that if a scientist has predominantly chemistry journals in their cited references, they are a chemist; and if one uses predominantly mathematical journals in their cited references, they are a mathematician. This approach—finding a dominating (or modal) value among journals in cited references—is much more fine-grained than linking scientists only to the journals they published in, which is a more prevalent approach in bibliometric studies.

However, we did not use a more detailed approach in which scientists change their disciplines over time and in which (in the Polish case) a 30-year period (1992–2021) was divided into five six-year periods, and 152,043 scientists were ascribed to their disciplines separately in each period (Kwiek & Roszka, 2024). The rationale behind our approach was to have an unequivocal discipline for each scientist in the sample. An alternative classification would be to use the Science-Metrix (2018) list of fields and subfields. However, as the major dimension was men versus women rather than fields, we do not expect the results to be substantially different. We could have also used different gender-detection tools (e.g., genderize.io), possibly with a slightly different gender composition of the sample (Kwiek & Szymula, 2026; Wais, 2016).

What other variables are not available in cross-national studies like ours—and, in contrast, are available in single-nation studies? Apart from biological age and nationally defined disciplines, academic positions and the dates of academic promotions are not available, including the date of PhD conferral. We do not know if or when the scientists were promoted to associate or full professorships, which could shed more light on gender promotion gaps (Suer, 2023), especially in the context of solid research productivity data. The only way to use proxies for academic positions is to produce groups (e.g., beginners, early-career, mid-career, and late-career scientists) based on academic age. Following this line of thinking, late-career scientists (at least 25 years of publishing experience) could be treated as equivalents of full professors (as in Kwiek and Szymula (2025b), who studied changes in productivity classes from a longitudinal perspective of late-career scientists or those with at least 25 years of publishing experience). We have promotion dates only for some countries (the Poland OPI data set, United States Academic Analytics data set, Norway Cristin data set, and Italy Ministry of Science data set). Promotion dates and the year of PhD conferral are not available in bibliometric data sets. Additionally, the bibliometric sources, that is, the two commercial data sets (Scopus and Web of Science), have their own limitations related to coverage, disciplinary focus, and languages. National publications in languages other than English are mostly excluded from these data sets. Consequently, the Scopus data set used in this study does not show all work published by survivors in science, even though the vast majority of literature in STEMM is published in English and indexed in Scopus (which is not the case for the humanities, which for this reason were removed from the sample and further analysis; for a more substantial discussion of the limitations of bibliometric studies and "measuring research," see Sugimoto and Larivière (2018, pp. 119–129) and Sugimoto and Larivière (2023, pp. 8–12)).

Despite the above limitations, we believe that this study contributes to two lines of research: global research on academic careers (Huang et al., 2020; Sugimoto & Larivière, 2023; Wang & Barabási, 2021) and global research on gender disparities in science (King et al., 2017; Larivière et al., 2013; Morgan et al., 2021). In both cases, we examine a highly selective segment of science, scientists working for more than two decades, men and women alike, extending our understanding of gender disparities in science. Our findings on women survivors highlight traditional assumptions on gender disparities in science in general from a new perspective: Our longitudinal study shows, using massive longitudinal empirical material, that almost all gender disparities (except for publishing productivity) do not hold for this segment of scientists.

Gender differences documented for larger populations of scientists, and to a large extent absent among long-term survivors, indicate that the structure of gender inequalities is strongly related to career stage. After more than two decades of uninterrupted presence in science, men and women survivors are similar in several major dimensions of academic work, but a substantial gender productivity gap exists.

The selectivity of the survivor population does not make the convergence of careers of men and women we observed a methodologically predetermined result. If long-term staying in science was leading to full homogenization of male and female careers, the disappearance of differences in all the dimensions we analyzed should be expected. But this is not the case. A gender productivity gap remains clearly visible after two decades and in the last five years of observation. At the same time, differences in international collaboration patterns, journal prestige, citation impact, and team size are small, absent, or—in some cases and in some disciplines—favor women. Our results do not indicate full homogenization of careers but selective, dimension-dependent convergence of careers between men and women among long-term survivors.

This distinction is important in the interpretation of gender inequality over the career course. Inequalities may be powerful at the levels of entering, staying, and exiting science, and at the same time they can be much weaker in some dimensions of academic work among late-career scientists. The absence of a particular gender gap conditional on survival does not mean its absence in the whole population of scientists. We can see where in scientific careers a particular inequality is empirically observable and where it ceases to be visible.

Gender inequality is powerfully dependent on career stage: Considerable inequalities are present in long-term survival in science, and among survivors, men and women become similar in some dimensions of academic work, but not in publishing productivity. Gender inequality in scientific careers may manifest itself in two analytically separate dimensions: inequality in who remains in science and inequality among those who remain in science.

A study of the minority of scientists provides insights about the academic profession as a whole—but after a certain point in time. Longitudinal studies introduce a temporal dimension to studies of gender inequality, one previously unavailable. Our cohort and longitudinal approach open the way to making gender comparisons of the "apples with apples" type (Nygaard et al., 2022): men and women starting their careers at the same moment and followed over time in the same disciplines. Our focus was on gender and disciplines rather than national science systems, and our data set of 38 countries provides a solid background and a substantial number of observations to revisit gender gaps in science.

Our findings show what was not previously apparent: There are large minority segments in science (in both STEMM and social science disciplines and across 38 OECD countries studied) where major traditional gender disparities, except for publishing productivity, do not exist. These findings have both theoretical implications for studies of gender inequality in science and practical implications for early versus late-career women in science. From a longitudinal perspective, after surviving in science for about 20 years, women tend not to differ from men surviving in science for the same period in major dimensions of academic work. Time matters in science, and longitudinal studies show its importance in career research. "Survivors in science," as a selective segment of science, provide an inspiring testing ground to study gender inequalities and their disappearance among late-career scientists.

## ACKNOWLEDGMENTS

We gratefully acknowledge the assistance of the International Center for the Studies of Research (ICSR) Lab, with particular gratitude to Kristy James and Alick Bird.

## AUTHOR CONTRIBUTIONS

Marek Kwiek: Conceptualization, Data curation, Formal analysis, Investigation, Methodology, Resources, Software, Validation, Writing—original draft, Writing—review & editing. Lukasz Szymula: Conceptualization, Data curation, Formal analysis, Investigation, Methodology, Software, Validation, Visualization, Writing—original draft, Writing—review & editing.

## COMPETING INTERESTS

The authors have no competing interests.

## FUNDING INFORMATION

This publication was co-financed from the state budget as part of the Minister of Education and Science's program "Science for Society II," project No. NdS-II/SP/0010/2023/01, with a grant amount of 750,000 PLN and a total project value of 750,000 PLN.

## DATA AVAILABILITY

We used data from Scopus, a proprietary scientometrics database. For legal reasons, data from Scopus received through collaboration with the ICSR Lab (Elsevier) cannot be made openly available.

## REFERENCES

Abramo, G., D'Angelo, C. A., & Di Costa, F. (2019). A gender analysis of top scientists' collaboration behavior: Evidence from Italy. *Scientometrics*, *120*, 405–418. https://doi.org/10.1007/s11192-019-03136-6

Abramo, G., D'Angelo, C. A., & Murgia, G. (2013). Gender differences in research collaboration. *Journal of Informetrics*, *7*(4), 811–822. https://doi.org/10.1016/j.joi.2013.07.002

Abramo, G., D'Angelo, C. A., & Rosati, F. (2015). Selection committees for academic recruitment: Does gender matter? *Research Evaluation*, *24*(4), 392–404. https://doi.org/10.1093/reseval/rvv019 **Q4**

Ackers, L. (2008). Internationalization, mobility, and metrics: A new form of indirect discrimination? *Minerva*, *46*, 411–435. https://doi.org/10.1007/s11024-008-9110-2

Aksnes, D. W., Piro, F. N., & Rørstad, K. (2019). Gender gaps in international research collaboration: A bibliometric approach. *Scientometrics*, *120*, 747–774. https://doi.org/10.1007/s11192-019-03155-3

Aksnes, D. W., Rørstad, K., Piro, F. N., & Sivertsen, G. (2011). Are female researchers less cited? A large-scale study of Norwegian researchers. *Journal of the American Society for Information Science and Technology*, *62*(4), 628–636. https://doi.org/10.1002/asi.21486

Barbezat, D. A., & Hughes, J. W. (2005). Salary structure effects and the gender pay gap in academia. *Research in Higher Education*, *46*(6), 621–640. https://doi.org/10.1007/s11162-004-4137-1

Bear, J. B., & Woolley, A. W. (2011). The role of gender in team collaboration and performance. *Interdisciplinary Science Reviews*, *36*(2), 146–153. https://doi.org/10.1179/030801811X13013181961473

Bozeman, B., Fay, D., & Slade, C. P. (2012). Research collaboration in universities and academic entrepreneurship: The-state-of-the-art. *The Journal of Technology Transfer*, *38*, 1–67. https://doi.org/10.1007/s10961-012-9281-8

Ceci, S. J., Ginther, D. K., Kahn, S., & Williams, W. M. (2014). Women in academic science: A changing landscape. *Psychological Science in the Public Interest*, *15*(3), 75–141. https://doi.org/10.1177/1529100614541236

Ceci, S. J., Kahn, S., & Williams, W. M. (2023). Exploring gender bias in six key domains of academic science: An adversarial collaboration. *Psychological Science in the Public Interest*, *24*(1), 15–73. https://doi.org/10.1177/15291006231163179, PubMed: 37098793

Chaoqun, N., Basson, I., Badia, G., Tufenkji, N., Sugimoto, C. R., & Larivière, V. (2025). Gender differences in submission behavior exacerbate publication disparities in elite journals. *eLife*, *12*, RP90049. https://doi.org/10.7554/eLife.90049.4

Clauset, A., Arbesman, S., & Larremore, D. B. (2015). Systematic inequality and hierarchy in faculty hiring networks. *Science Advances*, *1*(1), e1400005. https://doi.org/10.1126/sciadv.1400005, PubMed: 26601125

Cole, J. R. (1979). *Women in the scientific community*. New York, NY: Columbia University Press.

Cornelius, R., Constantinople, A., & Gray, J. (1988). The chilly climate: Fact or artifact? *The Journal of Higher Education*, *59*(5), 527–555. https://doi.org/10.1080/00221546.1988.11780211

Cruz-Castro, L., & Sanz-Menéndez, L. (2019). *Grant allocation disparities from a gender perspective: Literature review. Synthesis report*. GRANteD Project D.1.1. https://doi.org/10.20350/digitalCSIC/10548

Cummings, W. K., & Finkelstein, M. J. (2012). *Scholars in the changing American academy. New contexts, new rules and new roles*. Dordrecht: Springer. https://doi.org/10.1007/978-94-007-2730-4

Dabo-Niang, S., Esteban, M. J., Guillope, C., & Roy, M.-F. (2024). Aspects of the gender gap in mathematics. *EMS Journal*, *131*, 22–31. https://doi.org/10.4171/mag/178

Diezmann, C., & Grieshaber, S. (2019). *Women professors. Who makes it and how?* Singapore: Springer. https://doi.org/10.1007/978-981-13-3685-0

Ductor, L., Goyal, S., & Prummer, A. (2023). Gender and collaboration. *The Review of Economics and Statistics*, *105*(6), 1366–1378. https://doi.org/10.1162/rest_a_01113

Ehrenberg, R. G., Kasper, H., & Rees, D. I. (1991). Faculty turnover in American colleges and universities. *Economics of Education Review*, *10*(2), 99–110. https://doi.org/10.1016/0272-7757(91)90002-7

Elsevier. (2020). *The researcher journey through a gender lens*. Elsevier.

Feeney, M. K., & Bernal, M. (2010). Women in STEM networks: Who seeks advice and support from women scientists? *Scientometrics*, *85*, 767–790. https://doi.org/10.1007/s11192-010-0256-y

Fell, C. B., & König, C. J. (2016). Is there a gender difference in scientific collaboration? A scientometric examination of co-authorships among industrial-organizational psychologists. *Scientometrics*, *108*, 113–141. https://doi.org/10.1007/s11192-016-1967-5

Fowler, J. H., & Aksnes, D. W. (2007). Does self-citation pay? *Scientometrics*, *72*, 427–437. https://doi.org/10.1007/s11192-007-1777-2

Fox, M. F., & Mohapatra, S. (2007). Social-organizational characteristics of work and publication productivity among academic scientists in doctoral-granting departments. *The Journal of Higher Education*, *78*(5), 542–571. https://doi.org/10.1080/00221546.2007.11772329

Fox, M. F., & Nikivincze, I. (2021). Being highly prolific in academic science: Characteristics of individuals and their departments. *Higher Education*, *81*, 1237–1255. https://doi.org/10.1007/s10734-020-00609-z

Fox, M. F., Realff, M. L., Rueda, D. R., & Morn, J. (2017). International research collaboration among women engineers: Frequency and perceived barriers, by regions. *Journal of Technology Transfer*, *42*, 1292–1306. https://doi.org/10.1007/s10961-016-9512-5

Frehill, L. M., & Zippel, K. (2010). Gender and international collaborations of academic scientists and engineers: Findings from the survey of doctorate recipients, 2006. *Journal of the Washington Academy of Sciences*, *97*(1), 49–69.

Ghiasi, G., Mongeon, P., Sugimoto, C., & Larivière, V. (2018). Gender homophily in citations. *3rd International Conference on Science and Technology Indicators (STI 2018)* (pp. 1519–1525).

Glenn, N. D. (2005). *Cohort analysis*. Thousand Oaks, CA: Sage. https://doi.org/10.4135/9781412983662

Goastellec, G., & Vaira, M. (2017). Women's place in academia: Case studies of Italy and Switzerland. In H. Eggins (Ed.), *The changing role of women in higher education* (pp. 173–191). Cham: Springer. https://doi.org/10.1007/978-3-319-42436-1_9

Halevi, G. (2019). Bibliometric studies on gender disparities in science. In W. Glänzel, H. F. Moed, U. Schmoch, & M. Thelwall (Eds.), *Springer handbook of science and technology indicators* (pp. 563–580). Cham: Springer. https://doi.org/10.1007/978-3-030-02511-3_21

Heffner, A. G. (1979). Authorship recognition of subordinates in collaborative research. *Social Studies of Science*, *9*(3), 377–384. https://doi.org/10.1177/030631277900900305

Huang, J., Gates, A. J., Sinatra, R., & Barabási, A.-L. (2020). Historical comparison of gender inequality in scientific careers across countries and disciplines. *Proceedings of the National Academy of Sciences*, *117*(9), 4609–4616. https://doi.org/10.1073/pnas.1914221117, PubMed: 32071248

Hutson, S. R. (2006). Self-citation in archaeology: Age, gender, prestige, and the self. *Journal of Archaeological Method and Theory*, *13*, 1–18. https://doi.org/10.1007/s10816-006-9001-5

Ioannidis, J. P. A., Boyack, K. W., & Klavans, R. (2014). Estimates of the continuously publishing core in the scientific workforce. *PLOS ONE*, *9*(7), e101698. https://doi.org/10.1371/journal.pone.0101698, PubMed: 25007173

Kaminski, D., & Geisler, C. (2012). Survival analysis of faculty retention in science and engineering by gender. *Science*, *335*,

864–866. https://doi.org/10.1126/science.1214844, PubMed: 22344445

Kegen, N. V. (2013). Science networks in cutting-edge research institutions: Gender homophily and embeddedness in formal and informal networks. *Procedia—Social and Behavioral Sciences*, *79*, 62–81. https://doi.org/10.1016/j.sbspro.2013.05.057

Key, E., & Sumner, J. L. (2019). You research like a girl: Gendered research agendas and their implications. *PS: Political Science & Politics*, *52*(4), 663–668. https://doi.org/10.1017/S1049096519000945

King, M. M., Bergstrom, C. T., Correll, S. J., Jacquet, J., & West, J. D. (2017). Men set their own cites high: Gender and self-citation across fields and over time. *Socius*, *3*, 1–22. https://doi.org/10.1177/2378023117738903

Koffi, M. (2025). Innovative ideas and gender (in)equality. *American Economic Review*, *115*(7), 2207–2236. https://doi.org/10.1257/aer.20211811

Kwiek, M., & Roszka, W. (2021a). Gender disparities in international research collaboration: A large-scale bibliometric study of 25,000 university professors. *Journal of Economic Surveys*, *35*(5), 1344–1388. https://doi.org/10.1111/joes.12395

Kwiek, M., & Roszka, W. (2021b). Gender-based homophily in research: A large-scale study of man-woman collaboration. *Journal of Informetrics*, *15*(3), 101171. https://doi.org/10.1016/j.joi.2021.101171

Kwiek, M., & Roszka, W. (2022a). Academic vs. biological age in research on academic careers: A large-scale study with implications for scientifically developing systems. *Scientometrics*, *127*, 3543–3575. https://doi.org/10.1007/s11192-022-04363-0

Kwiek, M., & Roszka, W. (2022b). Are female scientists less inclined to publish alone? The gender solo research gap. *Scientometrics*, *127*, 1697–1735. https://doi.org/10.1007/s11192-022-04308-7

Kwiek, M., & Roszka, W. (2024). Top research performance in Poland over three decades: A multidimensional micro-data approach. *Journal of Informetrics*, *18*(4), 101595. https://doi.org/10.1016/j.joi.2024.101595

Kwiek, M., & Szymula, L. (2023). Young male and female scientists: A quantitative exploratory study of the changing demographics of the global scientific workforce. *Quantitative Science Studies*, *4*(4), 902–937. https://doi.org/10.1162/qss_a_00276

Kwiek, M., & Szymula, L. (2025a). Quantifying attrition in science: A cohort-based, longitudinal study of scientists in 38 OECD countries. *Higher Education*, *89*, 1465–1493. https://doi.org/10.1007/s10734-024-01284-0

Kwiek, M., & Szymula, L. (2025b). Quantifying lifetime productivity changes: A longitudinal study of 320,000 late-career scientists. *Quantitative Science Studies*, *6*, 1002–1038. https://doi.org/10.1162/QSS.a.16

Kwiek, M., & Szymula, L. (2026). Women in science: Measuring participation in Europe across disciplines, generations and over time. *Humanities and Social Sciences Communications*, *13*, 759. https://doi.org/10.1057/s41599-026-06912-x

Larivière, V., Ni, C., Gingras, Y., Cronin, B., & Sugimoto, C. R. (2013). Global gender disparities in science. *Nature*, *504*, 211–213. https://doi.org/10.1038/504211a

Larivière, V., Vignola-Gagné, E., Villeneuve, C., Gelinas, P., & Gingras, Y. (2011). Sex differences in research funding, productivity and impact: An analysis of Quebec university professors. *Scientometrics*, *87*, 483–498. https://doi.org/10.1007/s11192-011-0369-y

Leisyte, L., & Hosch-Dayican, B. (2017). Gender and academic work at a Dutch university. In H. Eggins (Ed.), *The changing role of women in higher education* (pp. 95–117). Cham: Springer. https://doi.org/10.1007/978-3-319-42436-1_5

Lerchenmueller, M., Hoisl, K., & Schmallenbach, L. (2019). Homophily, biased attention, and the gender gap in science. *Paper presented at DRUID19*, Copenhagen Business School, Copenhagen, Denmark, 19–21 June. https://doi.org/10.5465/AMBPP.2019.14784abstract

Maddi, A., Larivière, V., & Gingras, Y. (2019). Man-woman collaboration behaviors and scientific visibility: Does gender affect the academic impact in economics and management? *Proceedings of the 17th International Conference on Scientometrics & Informetrics* (pp. 1687–1697).

Madison, G., & Fahlman, P. (2020). Sex differences in the number of scientific publications and citations when attaining the rank of professor in Sweden. *Studies in Higher Education*, *46*(12), 2506–2527. https://doi.org/10.1080/03075079.2020.1723533

Maliniak, D., Powers, R., & Walter, B. F. (2013). The gender citation gap in international relations. *International Organization*, *67*(4), 889–922. https://doi.org/10.1017/S0020818313000209

Menard, S. (2002). *Longitudinal research*. Thousand Oaks, CA: Sage. https://doi.org/10.4135/9781412984867

Mihaljević-Brandt, H., Santamaria, L., & Tullney, M. (2016). The effect of gender in the publication patterns in mathematics. *PLOS ONE*, *11*(10), e0165367. https://doi.org/10.1371/journal.pone.0165367, PubMed: 27780266

Miller, J., & Chamberlin, M. (2000). Women are teachers, men are professors: A study of student perceptions. *Teaching Socioliology*, *28*(4), 283–298. https://doi.org/10.2307/1318580

Mishra, S., Fegley, B. D., Diesner, J., & Torvik, V. I. (2018). Self-citation is the hallmark of productive authors, of any gender. *PLOS ONE*, *13*(9), e0195773. https://doi.org/10.1371/journal.pone.0195773, PubMed: 30256792

Morgan, A. C., Way, S. F., Hoefer, M. J. D., Larremore, D. B., Galesic, M., & Clauset, A. (2021). The unequal impact of parenthood in academia. *Science Advances*, *7*(9), eabd1996. https://doi.org/10.1126/sciadv.abd1996

Nielsen, M. W. (2016). Gender inequality and research performance: Moving beyond individual-meritocratic explanations of academic advancement. *Studies in Higher Education*, *41*(11), 2044–2060. https://doi.org/10.1080/03075079.2015.1007945

Nygaard, L. P., Aksnes, D. W., & Piro, F. N. (2022). Identifying gender disparities in research performance: The importance of comparing apples with apples. *Higher Education*, *84*, 1127–1142. https://doi.org/10.1007/s10734-022-00820-0

Ployhart, R. E., & Vandenberg, R. J. (2010). Longitudinal research: The theory, design, and analysis of change. *Journal of Management*, *36*(1), 94–120. https://doi.org/10.1177/0149206309352110

Potthoff, M., & Zimmermann, F. (2017). Is there a gender-based fragmentation of communication science? An investigation of the reasons for the apparent gender homophily in citations. *Scientometrics*, *112*, 1047–1063. https://doi.org/10.1007/s11192-017-2392-0

Preston, A. E. (2004). *Leaving science*. New York, NY: Russell Sage Foundation.

Rivera, L. A. (2017). When two bodies are (not) a problem: Gender and relationship status discrimination in academic hiring. *American Sociological Review*, *82*, 1111–1138. https://doi.org/10.1177/0003122417739294

Rosser, V. J. (2004). Faculty members' intentions to leave: A national study on their work-life and satisfaction. *Research in Higher Education*, *45*, 285–309. https://doi.org/10.1023/B:RIHE.0000019591.74425.f1

Rowland, D. T. (2014). *Demographic methods and concepts*. Oxford: Oxford University Press.

Sarsons, H. (2017). Recognition for group work: Gender differences in academia. *American Economic Review*, *107*(5), 141–145. https://doi.org/10.1257/aer.p20171126

Sarsons, H., Gërxhani, K., Reuben, E., & Schram, A. (2020). Gender differences in recognition for group work. *Journal of Political Economy*, *129*(1), 101–147. https://doi.org/10.1086/711401

Science-Metrix. (2018). *Analytical support for bibliometrics indicators: Development of bibliometric indicators to measure women's contribution to scientific publications*. Final Report January 2018.

Singer, J. D., & Willett, J. B. (2003). *Applied longitudinal data analysis: Modeling change and event occurrence*. Oxford: Oxford University Press. https://doi.org/10.1093/acprof:oso/9780195152968.001.0001

Smart, J. C. (1990). A causal model of faculty turnover intentions. *Research in Higher Education*, *31*(5), 405–424. https://doi.org/10.1007/BF00992710

Spoon, K., LaBerge, N., Wapman, K. H., Zhang, S., Morgan, A. C., … Clauset, A. (2023). Gender and retention patterns among US faculty. *Science Advances*, *9*https://doi.org/10.1126/sciadv.adi2205, PubMed: 37862417

Süer, M. (2023). Are women in science less ambitious than men? Experimental evidence on the role of gender and STEM in promotion applications. *Rationality and Competition Discussion Paper Series 483* CRC TRR 190 Rationality and Competition.

Sugimoto, C., & Larivière, V. (2023). *Equity for women in science: Dismantling systemic barriers to advancement*. Cambridge, MA: Harvard University Press.

Sugimoto, C. R., & Larivière, V. (2018). *Measuring research. What everyone needs to know*. Oxford: Oxford University Press. https://doi.org/10.1093/wentk/9780190640118.001.0001

Tatarini, G., Gorodetskaya, O., & Vitali, A. (2024). Parenthood premium but fatherhood super-premium in academic productivity? A matter of partner's employment. *Community, Work, and Family*, 1–27. https://doi.org/10.1080/13668803.2024.2410433

Thelwall, M. (2020). Gender differences in citation impact for 27 fields and six English-speaking countries 1996–2014. *Quantitative Science Studies*, *1*(2), 599–617. https://doi.org/10.1162/qss_a_00038

Thelwall, M., Bailey, C., Tobin, C., & Bradshaw, N.-A. (2019). Gender differences in research areas, methods and topics: Can people and thing orientations explain the results? *Journal of Informetrics*, *13*(1), 149–169. https://doi.org/10.1016/j.joi.2018.12.002

Toutkoushian, R. K., & Bellas, M. L. (1999). Faculty time allocations and research productivity: Gender, race and family effects. *Review of Higher Education*, *22*(4), 367–390. https://doi.org/10.1353/rhe.1999.0014

Uhly, K. M., Visser, L. M., & Zippel, K. S. (2017). Gendered patterns in international research collaborations in academia. *Studies in Higher Education*, *42*(4), 760–782. https://doi.org/10.1080/03075079.2015.1072151

van den Besselaar, P., & Sandström, U. (2015). Early career grants, performance, and careers: A study on predictive validity of grant decisions. *Journal of Informetrics*, *9*(4), 826–838. https://doi.org/10.1016/j.joi.2015.07.011

van den Besselaar, P., & Sandström, U. (2016). Gender differences in research performance and its impact on careers: A longitudinal case study. *Scientometrics*, *106*, 143–162. https://doi.org/10.1007/s11192-015-1775-3, PubMed: 26798162

van den Besselaar, P., & Sandström, U. (2017). Vicious circles of gender bias, lower positions, and lower performance: Gender differences in scholarly productivity and impact. *PLOS ONE*, *12*(8), e0183301. https://doi.org/10.1371/journal.pone.0183301, PubMed: 28841666

van den Brink, M., & Benschop, Y. (2013). Gender in academic networking: The role of gatekeepers in professorial recruitment. *Journal of Management Studies*, *51*(3), 460–492. https://doi.org/10.1111/joms.12060

Wachter, K. W. (2014). *Essential demographic methods*. Cambridge, MA: Harvard University Press. https://doi.org/10.4159/9780674369757

Wais, K. (2016). Gender prediction methods based on first names with genderizeR. *The R Journal*, *8*(1), 17–37. https://doi.org/10.32614/RJ-2016-002

Wang, D., & Barabási, A.-L. (2021). *The science of science*. Cambridge: Cambridge University Press. https://doi.org/10.1017/9781108610834

Ward, M. E., & Sloane, P. J. (2000). Non-pecuniary advantages versus pecuniary disadvantages: Job satisfaction among male and female academics in Scottish universities. *Scottish Journal of Political Economy*, *47*(3), 273–303. https://doi.org/10.1111/1467-9485.00163

Weisshaar, K. (2017). Publish and perish? An assessment of gender gaps in promotion to tenure in academia. *Social Forces*, *96*(2), 529–560. https://doi.org/10.1093/sf/sox052

White-Lewis, D. K., O'Meara, K., Mathews, K., & Havey, N. (2023). Leaving the institution or leaving the academy? Analyzing the factors that faculty weigh in actual departure decisions. *Research in Higher Education*, *64*, 473–494. https://doi.org/10.1007/s11162-022-09712-9

Wohrer, V. (2014). To stay or to go? Narratives of early-stage sociologists about persisting in academia. *Higher Education Policy*, *27*, 469–487. https://doi.org/10.1057/hep.2014.22

Xu, Y. J. (2008). Gender disparity in STEM disciplines: A study of faculty attrition and turnover intentions. *Research in Higher Education*, *49*, 607–624. https://doi.org/10.1007/s11162-008-9097-4

Zhou, Y., & Volkwein, J. F. (2004). Examining the influence on faculty departure intentions: A comparison of tenured versus nontenured faculty at research universities using NSOPF-99. *Research in Higher Education*, *45*(2), 139–176. https://doi.org/10.1023/B:RIHE.0000015693.38603.4c